\documentclass[%
reprint,
superscriptaddress,
amsmath,amssymb,
prstab,
]{revtex4-1}

\usepackage{graphicx}% Include figure files
\usepackage{stackengine}
\usepackage{dcolumn}% Align table columns on decimal point
\usepackage{bm}% bold math
\usepackage{overpic}% super impose on the picture
\usepackage{pict2e}
\usepackage{tikz}
\usepackage{natbib}
\usepackage[nottoc,numbib]{tocbibind}
\usepackage{amsmath}
\usepackage{gensymb}
\usepackage{tikz,xcolor}

\usepackage[%
  colorlinks=true,
  urlcolor=blue,
  linkcolor=blue,
  citecolor=blue
]{hyperref}

\definecolor{lime}{HTML}{A6CE39}
\DeclareRobustCommand{\orcidicon}{%
	\begin{tikzpicture}
	\draw[lime, fill=lime] (0,0) 
	circle [radius=0.16] 
	node[white] {{\fontfamily{qag}\selectfont \tiny ID}};
	\draw[white, fill=white] (-0.0625,0.095) 
	circle [radius=0.007];
	\end{tikzpicture}
	\hspace{-2mm}
}

\foreach \x in {A, ..., Z}{%
	\expandafter\xdef\csname orcid\x\endcsname{\noexpand\href{https://orcid.org/\csname orcidauthor\x\endcsname}{\noexpand\orcidicon}}
}
\begin{document}

\preprint{AIP/123-QED}

\title{Pushing the Capture Limit of Thermionic Gun Linacs}
%\title{High Capture Efficiency Cavity Design for Radiation Therapy Treatment Applications}
%\title{A Compact High Capture Efficiency Radiotherapy Linac Design \\Using Low Voltage Velocity Bunching}
%\\with Forced Linebreak\footnote{Error!}}% Force line breaks with 
%\thanks{Footnote to title of article.}
\newcommand{\orcidauthorA}{0000-0002-5903-8930} % Sadiq; Add \orcidA{} behind the author's name
\newcommand{\orcidauthorB}{0000-0003-1040-7640} % Alex; Add \orcidA{} behind the author's name
\newcommand{\orcidauthorC}{0000-0003-2185-2897} % Julian; Add \orcidA{} behind the author's name
\newcommand{\orcidauthorD}{0000-0001-6346-5989} % Rob; Add \orcidA{} behind the author's name
\newcommand{\orcidauthorE}{0000-0003-0813-3521} % Boris; Add \orcidA{} behind the author's name
\newcommand{\orcidauthorF}{0000-0001-5727-3526} % Deepa; Add \orcidA{} behind the author's name
\newcommand{\orcidauthorG}{0000-0001-8288-1216} % Graeme; Add \orcidA{} behind the author's name

\author{Sadiq \surname{Setiniyaz}\orcidA{}}
\email{s.saitiniyazi@lancaster.ac.uk}
\affiliation{Engineering Department, Lancaster University, LA1 4YW, UK}
\affiliation{Cockcroft Institute, Daresbury Laboratory, Warrington, WA4 4AD, UK}

\author{Alejandro \surname{Castilla}\orcidB{}}
\affiliation{Engineering Department, Lancaster University, LA1 4YW, UK}
\affiliation{Cockcroft Institute, Daresbury Laboratory, Warrington, WA4 4AD, UK}

\author{Julian \surname{McKenzie}\orcidC{}}
\affiliation{Cockcroft Institute, Daresbury Laboratory, Warrington, WA4 4AD, UK}
\affiliation{ASTeC, STFC Daresbury Laboratory, Warrington, WA4 4AD, UK}

\author{Robert \surname{Apsimon}\orcidD{}}
\affiliation{Engineering Department, Lancaster University, LA1 4YW, UK}
\affiliation{Cockcroft Institute, Daresbury Laboratory, Warrington, WA4 4AD, UK}

\author{Boris \surname{Militsyn}\orcidE{}}
\affiliation{Cockcroft Institute, Daresbury Laboratory, Warrington, WA4 4AD, UK}
\affiliation{ASTeC, STFC Daresbury Laboratory, Warrington, WA4 4AD, UK}

\author{Deepa \surname{Angal-Kalinin}\orcidF{}}
\affiliation{Cockcroft Institute, Daresbury Laboratory, Warrington, WA4 4AD, UK}
\affiliation{ASTeC, STFC Daresbury Laboratory, Warrington, WA4 4AD, UK}

\author{Graeme \surname{Burt}\orcidG{}}
\affiliation{Engineering Department, Lancaster University, LA1 4YW, UK}
\affiliation{Cockcroft Institute, Daresbury Laboratory, Warrington, WA4 4AD, UK}

%\thanks{Fax: +82-42-866-6150}

\date{\today}% It is always \today, today,
% but any date may be explicitly specified

% I have been reviewing the paper on overleaf and I think it needs a complete restructure. It gives results of concepts that aren’t introduced till later in the paper and is hard to follow unless you read it twice as it jumps around with different codes and field maps. I attach the structure I think we need

%1. Introduction: commercial linacs, capture of e- in linacs, advantages of higher capture

%2. ASTRA sims of existing linacs (don’t show 1D tracking yet as its not introduced)

%3. New 1D code for analysis of capture, describe the code, how it works, reference klystron arrival time functions and applegate and why they are useful for linacs, limitations with space charge

%4. Apply new code to existing linacs, discuss issues

%5. Velocity bunching in linacs, how it is affected by beam energy, capture versus gradient

%6. Using new code to optimise arrival time functions (can we do for 6 MeV linac rather than 8 MeV for consistency?) suing sinusoidal field maps only

%7. Proving in ASTRA with generic sinusoidal field maps

%8. RF cavity design (6 MeV linac) : I am working on this now but may need a week or so to fully develop theory

%9. 1D and ASTRA sims with real cavity (a few different field maps from RF cavity with say different cell lengths)

%10. Possible something on trans. acceptance from the gun ???

% Keep your message clear; Create a logical framework; State your case with confidence; Beware the curse of ‘zombie nouns’; Prune that purple prose; Aim for a wide audience; 

\begin{abstract}	
Although accelerator technology has matured sufficiently, state-of-the-art X-ray linacs for radiotherapy and cargo-scanning capture merely $30-50\%$ of the electrons from a thermionic cathode, requiring a higher cathode current and leaving uncaptured electrons to cause problems such as back bombardment on the cathode leading to a shortening of cathode life. Any solution to increase capture should be effective, simple, reliable, compact, and low cost in order to be adopted by the industry. To address this, we present the design of a 6 MeV high capture efficiency S-band electron linac that captures 90$\%$ of the initial DC beam. This linac does not require any extra parts that would increase the cost as the high efficiency is achieved via a low-field amplitude in the first bunching cell to decrease the number of backstreaming electrons, to velocity bunch the electron beam, and recapture backstreaming electrons. Under the low field amplitude, any electrons launched at decelerating phases travel backward with low speeds, thus most of them can catch the next RF cycle, and get re-accelerated/recaptured. As the electron speed is low, the cell length is also shorter than existing linacs. Such a short field is achieved by the use of asymmetric cells with differential coupling to the side-coupled cells. Our novel design has implications for all commercial high current thermionic gun linacs for increasing beam current and increasing cathode lifetime. 
%212 words 1,530 characters

	%Focus on the physics of the beam as well as the design for radiotherapy. The focus should be on the development of high capture commercial linacs.
\end{abstract}
%Valid PACS numbers may be entered using the \verb+\pacs{#1}+ command.
% Classification Scheme.
%29.20.−c Accelerators
%29.20.Ej Linear accelerators 
%41.75.Fr Electron and positron beams 
%87.56.bd Accelerators (for accelerators used in biological and medical physics)
%29.27.-a Beams in particle accelerators 

 % Use show keys class option if keyword % display desired
\maketitle
\section{INTRODUCTION}
Commercial linear accelerators have a wide range of applications with the main application being as a MeV-level X-ray source for the medical field of radiotherapy~\cite{MARTINS201478}. Radiotherapy is a key component in the treatment of many cancers, but a major barrier to scaling up the number of commercial linacs is the maintenance cost, both in terms of manpower and the financial cost, due to the regular replacement of components with a limited lifetime compared to the operating life of the machine. Increasing the mean time between failures of these components could significantly decrease the maintenance cost as well as reducing the amount of time the machine is down for repair. \cite{ClinicalOncology} One such component is the cathode in the thermionic electron gun~\cite{KORENEV2004537}. 

The linacs utilized for X-ray radiotherapy and cargo-scanning typically use thermionic cathodes that generate long macro-pulses of several microseconds~\cite{KarzmarkBook, Hernandez-Garcia2008, Hawkes2017}. Such macro-pulses cover many RF cycles and are therefore regarded as quasi-DC. While half the electrons enter the cavity at accelerating phases of the RF, the other half enters the cavity at decelerating RF phases, and get accelerated in the opposite direction, with most traveling backward thus end up hitting the cathode. This phenomenon is known as the Back Bombardment (BB)~\cite{MCKEE1991386, Bakr2011, Huang1991, KII2001, Kowalczyk2014, Kii2002, Edelen2014,MCKEE1990716,Borland:1991id} effect. The back-streaming electrons deposit their Kinetic Energy (KE) to the cathode and heat it up, which causes the cathode to generate more electrons and increase the beam current. As the beam current increases, more RF energy is taken away by the beam, and the cavity voltage drops, which causes the final beam energy to become unstable. In practice, the beam current and pulse length are limited to mitigate BB. BB shortens the cathode lifetime by heating it up so it is poisoned easily and degrades faster~\cite{Capece2015,Borland:1991id,Whelan2016}. 

Some work has been done to reduce the BB effect by external magnetic fields~\cite{MCKEE1991386, Huang1991}; hollow cathodes~\cite{Huang1991, Kii2002}; cooling~\cite{Kowalczyk2014}; and improvements to the cathode material~\cite{Bakr2018}. These methods would add extra cost and complicate the system. Here, we introduce a novel method to directly lower BB by increasing the capture efficiency and consequently reducing the number of electrons traveling backward. The capture efficiency is the percentage of electrons that reach the exit of the linac from those emitted from the cathode. In our design, we increased the capture efficiency from the $30 - 50\%$ of most standard existing DC thermionic gun linacs to over 90$\%$. Of the 10$\%$ uncaptured electrons, around 3$\%$ are lost by hitting the cavity aperture, and 7$\%$ are lost by backstreaming. In existing standard thermionic gun linacs, aperture loss and backstreaming loss are at least $15\%$ and 36$\%$, respectively. Thus, in our design BB electrons are reduced by more than 80$\%$ for a given gun current. 
The cathode lifetime can be further increased as due to the high capture efficiency, less current needs to be extracted from the cathode to supply the same amount of current to the target after acceleration. Consequently, the cathode can operate at a lower current density that will slow down wear out and increase lifetime~\cite{Palluel1980} as well as further halving the number of BB electrons. As the cathode current density is lowered, the cathode temperature can be lowered as well~\cite{nottingham1956thermionic}, which again increases the lifetime by abating the aforementioned poisoning. 

Another advantage of high capture efficiency is it reduces unwanted harmful radiation generated by electrons hitting the cavity walls. Electrons hitting the cavity walls will create X-rays and require shielding around the cavity~\cite{Swanson1979}, and/or the use of solenoid magnets to focus these electrons~\cite{KarzmarkBook}. This will increase gantry size, complicate the system, and eventually increase the cost. As the aperture loss is decreased by more than 90$\%$ in our cavity design, such a linac requires significantly less shielding. It also does not require a solenoid, although the addition of one would further reduce aperture loss, thus increasing capture. 

While other high capture efficiency designs are reported in the literature, they tend to be long and need a large gantry. The associated high manufacturing and maintenance cost would render them unfavorable options, especially for radiotherapy. In our design, the linac is a $\pi$/2-mode standing wave (SW) side-coupled accelerating structure, similar to the structures in Refs.~\cite{Knapp1968,Victor,wangler2008rf}, with a length shorter than 30~cm and beam energy of 6~MeV. This is achieved by having a step in the gradient between the first and second cell, allowing a low field capture cell, to allow low energy velocity bunching, before being captured and accelerated in a higher gradient section.

%There are other high capture but uncompact linac designs. A captures the efficiency of 60$\%$ design was reported by employing a long bunching cavity, low field gradient, and focusing solenoid for L-band linac, which required cavity to be 14 meters long. Higher efficiency of 90$\%$ S-band linac design reported that utilizes early cells as bunching cells, which also has low field gradient and focusing coils~\cite{Liu2006}. Because it is a constant gradient accelerating structure, the field is kept low in both accelerating and bunching cells, which required a total of 59 cells and a 2~m long cavity to reach 10~MeV energy.
%Alimov $et al.$~\cite{ Alimov2002} manufactured a 0.8~m long structure experimental 

\section{Capture Efficiency of Existing Commercial Linacs}
\label{sec:TradMedLin}
%2. ASTRA sims of existing linacs (don’t show 1D tracking yet as its not introduced)

In existing commercial linacs, the first cavity cell is roughly half the length of the other cells, so that in-phase electrons gain the maximum acceleration from the RF wave~\cite{KarzmarkBook}, while increasing the number of electrons that receive this acceleration. The accelerating gradient is constant along the standing wave cavity and ranges from 10$-$25 MeV/m depending on the design. To match the phase velocity to the electron velocity, the cell length is a function of velocity, $L=\frac{\beta\lambda}{2}$, where $\beta$ is the ratio of particle velocity to speed of light, and $\lambda$ is the RF wavelength. More than 50$\%$ of the beam is lost inside linacs as the linac can only capture electrons at certain phases.

The Varian Clinac-1800 accelerator captures phases between $-110^\circ$ and $20^\circ$, which is capture efficiency of 36$\%$~\cite{KarzmarkBook}. A study by Aubin $et$ $al.$~\cite{Aubin} reported 37$\pm$2$\%$ capture efficiency measured from a Varian 600C linac. Their simulation predicted that the capture efficiency can be increased up to 45$\%$ with a converging electron beam. Baillie $et$ $al.$~\cite{Baillie2017DesignAS} designed a variable-energy linac and reported capturing 133.7~mA out of 373~mA gun current, which is about 36$\%$ efficiency. 

\begin{figure}
	{\scalebox{0.4} [0.4]{\includegraphics{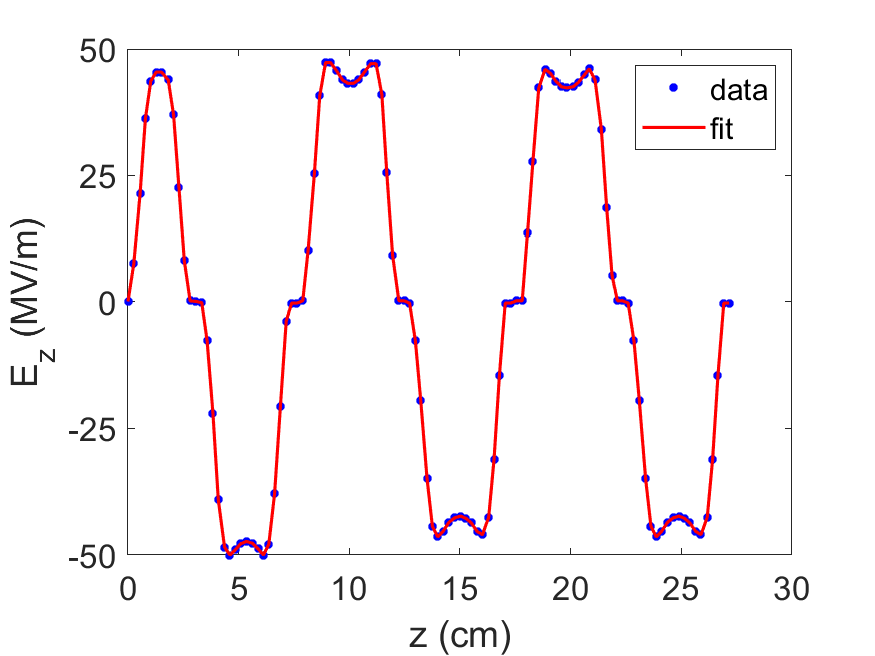}}}
	\caption{Longitudinal E-field profile of linac from Ref.~\cite{RIMJAEM2017233}.}
	\label{fig:TradMedLinacEz}
\end{figure}

ASTRA (A Space Charge Tracking Algorithm) is a 2.5D tracking code with space charge~\cite{ASTRA,Poplau:2006za}. ASTRA simulations were performed to study the capture and loss of an existing S-band linac structure with 6 cells by tracking electrons through an E-field map shown in Fig.~\ref{fig:TradMedLinacEz}~\cite{RIMJAEM2017233}. The blue dots are measurements made by the bead-pull technique, and the red line is the fit. Initial DC beams with 9.5, 15, and 25~keV Kinetic Energy (KE), and 100~mA current were simulated, with the results given in Table.~\ref{tab:TradMedCapture}. The ASTRA simulations agree well with the prediction of Baillie $et$ $al.$~\cite{Aubin}. When the E$_z$-field profile is scaled up so at its max E$_{z, max} = 50$~MV/m the final beam energy at the target (located at the exit of the linac) is 5.8~MeV.

%\textcolor{red}{start here 2021-02-05}
\begin{table}
	\caption{Simulated capture efficiency and loss of commercial linac.}
	\begin{ruledtabular}
		\begin{tabular}{lcccc}
			& \multicolumn{3}{c}{Initial KE} \\ \cline{2-4}
			 		&  	{9.5~keV} &  {15~keV}  & {25~keV}    \\				 			
			\hline
%			capture 			& 1D  	    & 43.3$\%$ &    46.1$\%$ &   48.6$\%$ \\
			Capture ($\%$)			 	& 41.6 &    45.4 &   48.0 \\
			Back-streaming loss ($\%$)  & 34.6&    35.5 &   36.6 \\		
			Aperture loss ($\%$) 	  	& 23.8 &    19.1 &   15.4 \\
%			beam size (mm) 	            & 1.1  &  1.2 &  1.3  \\
%			beam divergence (mrad) 	    &  &   &    \\
%			RMS energy spread (mrad) 	&  &   &    \\
		\end{tabular}
	\end{ruledtabular}
	\label{tab:TradMedCapture}
\end{table}

Literature reviews and our own simulations confirm that the capture efficiency of existing commercial linacs is close to 50$\%$ at best. 37$\%$ of this is the back-streaming electron loss and 15$\%$ are the electrons lost on the aperture. Existing commercial linacs in operation capture only $36-37\%$ of electrons, which indicates higher back-streaming loss and aperture loss. A major cause of this loss is that the accelerating/decelerating voltage of the first cell is much larger than the electron beam energy, causing almost half the electrons to be lost immediately. Significant improvement can be achieved if the gradient and lengths of the first few cells could be optimized.

\section{1D tracking code for cavity optimization}

%3. New 1D code for analysis of capture, describe the code, how it works, reference klystron arrival time functions and applegate and why they are useful for linacs, limitations with space charge
% \textcolor{red}{reference klystron arrival time functions} \textcolor{blue}{We don't use arrival time function. In our 1D code, time is the variable.}
\subsection{1D tracking code}

A 1D tracking code was developed to optimize the length and E-field amplitude of each cell by assessing the exit/arrival phases and energies of electrons at each cell as a function of the launch phase. This is a method originally proposed for the design of electron bunching for high-efficiency klystrons~\cite{Igor2015}. 
This code doesn't include space charge and only used for a fast and efficient scan of parameters, and that the optimized result from this initial optimization will be verified and further optimized by ASTRA simulations that include transverse plane and space charge effects.
The exit phase is the RF phase in which an electron exits a cell. The launch phase is the RF phase an electron enters the first cell. The aim is to maximize the range of launch phases which result in the same exit phase after the first few cells. The code needs as inputs: the E-field profile, RF frequency, particle charge, mass, and initial KE. Users also need to specify the maximum tracking time $T_{max}$ and number of time steps $N_{steps}$. The code computes the distance traveled by the particle $dz$ during the time interval of $dt = T_{max}/N_{steps}$. The velocity of the particle is obtained by using its KE. The change in the KE during this time interval is 
\begin{equation}
    d(\text{KE}) = E(z,t)dz
    \label{eq:dKE}
\end{equation}
\noindent where $E(z,t)$ is the longitudinal electric field seen by the particle at time $t$ and location $z$, which can be given as
\begin{equation}
    E(z,t) = E_{z}cos(2 \pi f t + \phi_{launch})
    \label{eq:Ezt}
\end{equation}
\noindent where $E_{z}$ is the longitudinal electric field profile, $f$ is the cavity frequency, and $\phi_{launch}$ is the launch phase. The electrons are launched at a fixed phase/time interval and tracked in the longitudinal direction until they exit from either end of the field profile. Electrons are launched at a range of phases to cover at least one full RF cycle. If an electron exits from the beginning or end of the field profile, it will be counted as lost or captured respectively. Electrons that do not reach the exit within the maximum tracking time are also counted as lost. Presently, the code neglects space charge and so has only be used for initial parameter scans, as its speed and advanced methods of optimization have provided approximate global optimum values. 

For optimization purposes, the code can be run in a scan mode, in which it varies the lengths and field amplitudes of each cell. An output file is generated for each cell length and field amplitude, which contains information such as the particle launch phase, exit phase, exit time, exit KE, and flag for whether the particle is lost or captured. The code can also be run in tracking mode, where we can investigate the beam dynamics for specific field settings. 

\subsection{Beam dynamics study of existing commercial linacs with 1D code}
%4. Apply new code to existing linacs, discuss issues

\begin{figure}
	\begin{tabular}{cc}
		\def\stackalignment{l}
		\topinset{\bfseries(a)}{\includegraphics[width=3in]{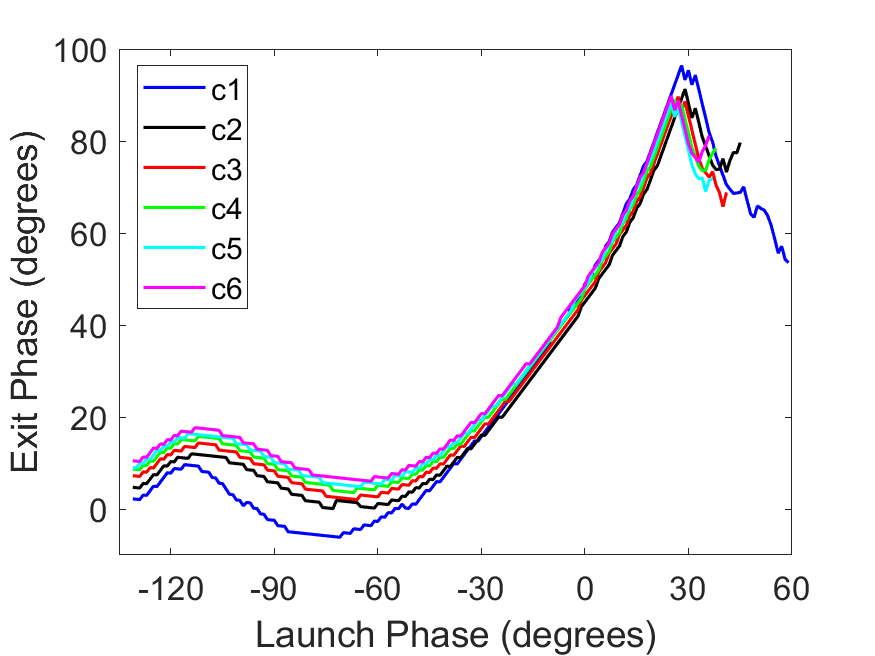}}{0.2in}{2.5in} \\
		\def\stackalignment{l}
		\topinset{\bfseries(b)}{\includegraphics[width=3in]{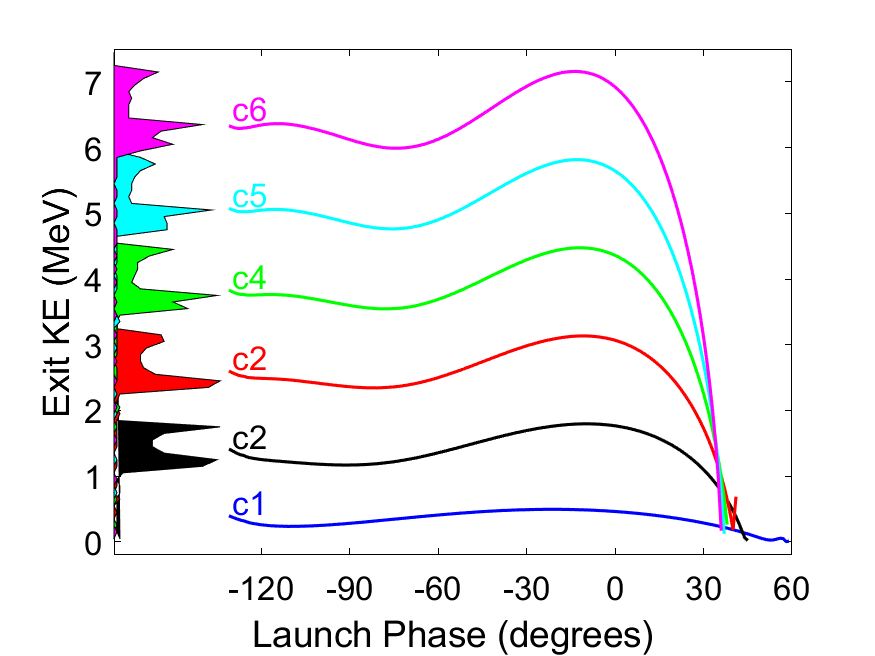}}{0.2in}{2.5in} \\		
	\end{tabular}
	\caption{Results of 1D tracking code simulations of a commercial linac with 25~keV initial electrons. Exit phase (a) and exit KE (b) vs launch phase for cells c1 through c6 of the linac. The KE energy in (b) is projected to the y-axis to show the KE spectrum in arbitrary units at the corresponding cell exits.}
	\label{fig:TradMedLinacExitPhase}
\end{figure}

We have studied the performance and beam dynamics of a commercial linac with the field profile of Fig.~\ref{fig:TradMedLinacEz} using the 1D code. The tracking results are shown in Fig.~\ref{fig:TradMedLinacExitPhase}. The exit phase, and KE, are given for all captured phases. The launched electrons covered a full RF cycle, i.e. 360$^\circ$, but only those launched within a phase range of $-131^\circ$ and 59$^\circ$ successfully exited the first cell. Electrons are further lost in later cells, and only those launched at phases between $-131^\circ$ and 36$^\circ$ successfully exited the last cell. This gives a capture efficiency of $167^\circ/360^\circ~=~46.4\%$. This is similar to the results of ASTRA simulations shown in section~\ref{sec:TradMedLin}. In Fig.~\ref{fig:TradMedLinacExitPhase} (a), we observe a slight narrowing of the exit phase range as electrons travel through each cell, which indicates bunching. In Fig.~\ref{fig:TradMedLinacExitPhase} (b), we can see cell by cell energy gain. It takes 6 cells for most electrons to reach 6~MeV. The electrons launched around $[$40$^\circ$,$- 60^\circ]$ phase range have very low energy. All this information is useful to understand the beam dynamics of the linac.

%Note that the zero launch phase corresponds to the peak of the accelerating field.

\begin{figure}
	{\scalebox{0.5} [0.5]{\includegraphics{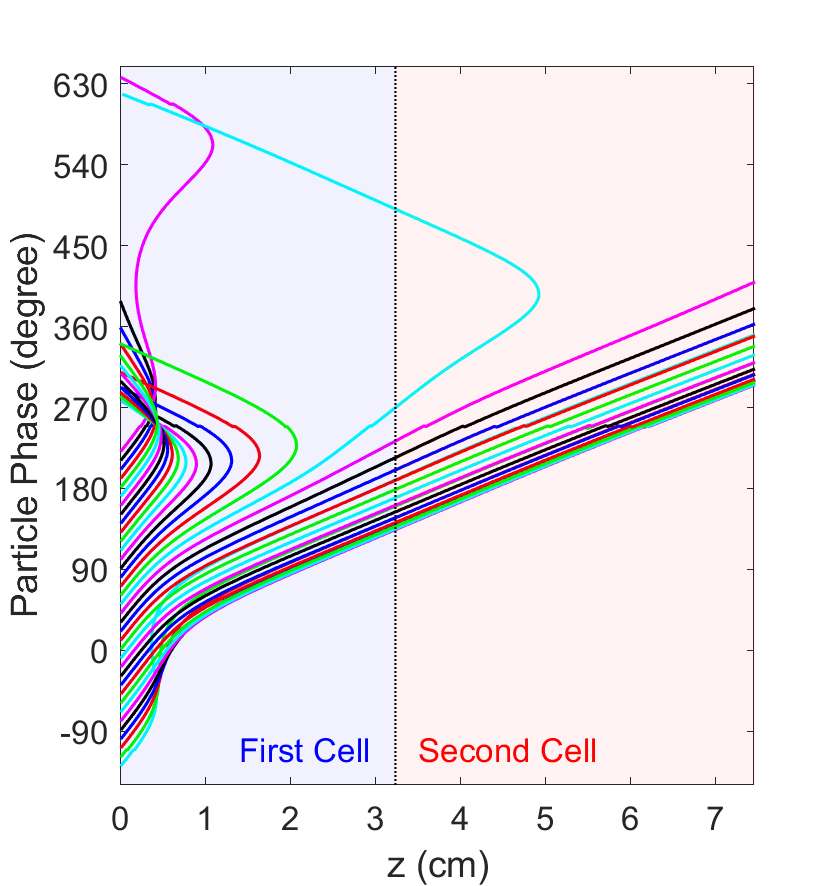}}}
	\caption{Applegate diagram of an existing standard commercial linac.}
	\label{fig:TradMedLinac}
\end{figure}

The 1D code can also generate an Applegate diagram, as shown in Fig.~\ref{fig:TradMedLinac}. Applegate diagrams are used in klystron design to show how electrons bunch as they travel along the cavity and are useful here to study the dynamics of bunching. The x-axis is the position of the particle inside the cavity and the y-axis is its phase. If the particle's position increases or decreases as the phase increases, then the particle is traveling forwards or backward respectively. If an electron travels backward and passes z~=~0, then this particle is lost. The range of phases occupied by the electrons indicates the amount of bunching. If the phase range gets smaller as the position increases, then electrons are coming closer together and bunching. The diagram in Fig.~\ref{fig:TradMedLinac} shows: that bunching mostly happened in the first cell; how far an electron travels before it is lost; and which phases are captured. From electrons emitted across the full 360$^\circ$ phase range, about half are launched at accelerating phases. The other half launched decelerating phases, travel backward, and thus become back-streaming electrons. 
As the field amplitude is high in the first cell, most of the back-streaming electrons cannot catch the next RF cycle, will continue traveling backward, and finally will be lost by hitting the cathode, other parts of the gun, and beam pipe. This causes heating and damage to the cathode. 

\section{Velocity bunching and recapture}

%5. Velocity bunching in linacs, how it is affected by beam energy, capture versus gradient 
%6. Using new code to optimise arrival time functions (can we do for 6 MeV linac rather than 8 MeV for consistency?) using sinusoidal fieldmaps only

In commercial accelerators, the electron source is a thermionic cathode that generates a continuous electron beam, of duration equal to applied RF pulse length. This is usually several microseconds long. The electron macro-pulse splits into many electron bunches under the RF field as some electrons are captured by accelerating phases and some are lost. For S-band, the RF wavelength is 333~ps long and so bunches are repeated every 333~ps. The captured electrons experience different acceleration and some travel faster than the others. In Fig.~\ref{fig:TradMedLinac}, we see early electrons (i.e. electrons from $-130^\circ$ to $-70^\circ$ phases) travel slower, and later, faster electrons catch up with them, which leads to an overall shortening of the bunch length. This process is called velocity bunching.

Velocity bunching takes place in the early cells of accelerators while the electron energy is sub-relativistic, and the bunch has a significant velocity gradient. As the beam energy increases, the electrons in the bunch have similar velocity, therefore velocity bunching cannot occur. The field gradients in the early cells are important as they determine the amount of bunching. A too high or low field gradient results in sub-optimal bunch length, and particle loss. 

In existing commercial linacs, the field in the first cell is too high for optimal bunching, as half of the phases are decelerated quickly and hit the cathode. Therefore, lowering the field amplitude in the first cell is the key to optimizing bunching and capture. Having a low field reduces the acceleration in the first cell, hence the relativistic $\beta$ is smaller. This implies the cell length should be shorter. However, we can make the cell longer than the one expected from the beta so that electrons that do not escape in the first RF cycle can be re-accelerated in the second RF cycle. They will then arrive at the second cell at the correct phase to be accelerated. In other words, the first cell should be longer than the synchronous beta-cell length to give additional delay for the recaptured electrons so they enter the second cell at an accelerating phase. This will cause them to form a bunch with electrons from the second RF cycle. The electrons that enter the second cell are not fully relativistic yet hence the length of the second cell needs to be shorter than that of later cells, so electrons enter the third cell at the right phase for acceleration. Electrons that exit the second cell are fully relativistic, therefore later cells need no adjustment for the beta.

The first scan in 1D code is coarse as we want to find a global optimum initially, in which four parameters (field length and amplitude of the first 2 cells) are scanned simultaneously. Once the global optimums are found, we performed fine scans near them in 1D code. Lastly, we performed even finer scans in ASTRA to fine-tune parameters. By the end of the fine-tuning, the capture efficiency changes by less than 0.1$\%$ between parameters around the optimums.
%L$_{c1}$ from 14 to 46~mm and amplitude E$_{z,max,c1}$ from 5 to 50~MV/m; and the second cell's length L$_{c2}$ from 14 to 82~mm and amplitude E$_{z,max,c2}$
\subsection{Optimization by using 1D code}

The optimization was performed by conducting a 4D grid scan, where four parameters of the first 2 cells are scanned. A field profile similar to the one in Fig.~\ref{fig:TradMedLinacEz} was used as the starting point in optimization. The initial KE used in the scan was 25~keV. The ranges scanned over for each parameter are: the first cell's length L$_{c1}$ from 10 to 46~mm and amplitude E$_{z,max,c1}$ from 4 to 50~MV/m; and the second cell's length L$_{c2}$ from 10 to 82~mm and amplitude E$_{z,max,c2}$ from 30 to 100~MV/m. 

\begin{figure*}
	\begin{tabular}{ccc}
		\def\stackalignment{l}	\topinset{\bfseries(a)}{\includegraphics[width=2in]{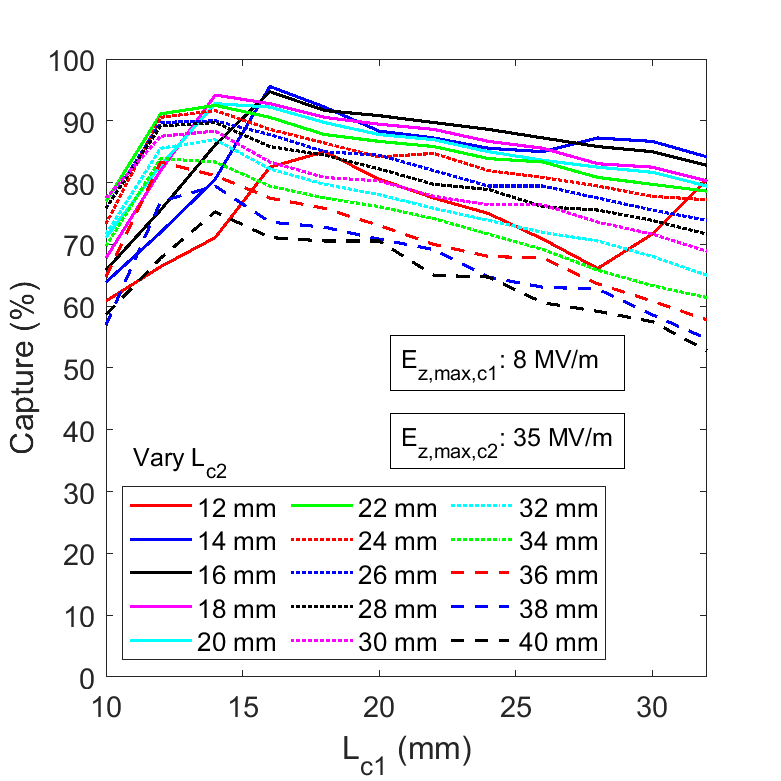}}{0.145in}{1.50in} & 	
		\def\stackalignment{l} 	\topinset{\bfseries(b)}{\includegraphics[width=2in]{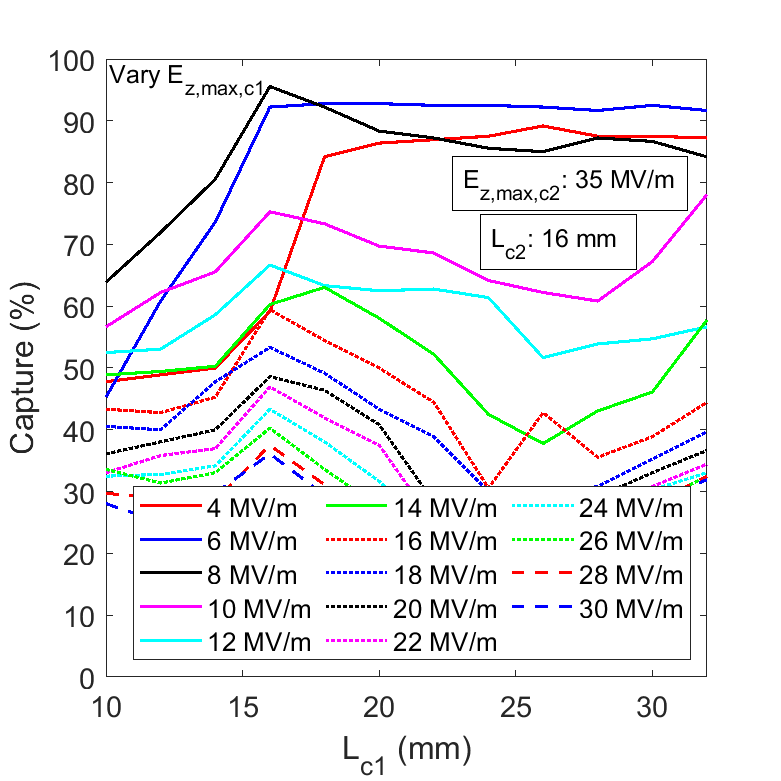}}{0.145in}{1.50in} &
		\def\stackalignment{l}  \topinset{\bfseries(c)}{\includegraphics[width=2in]{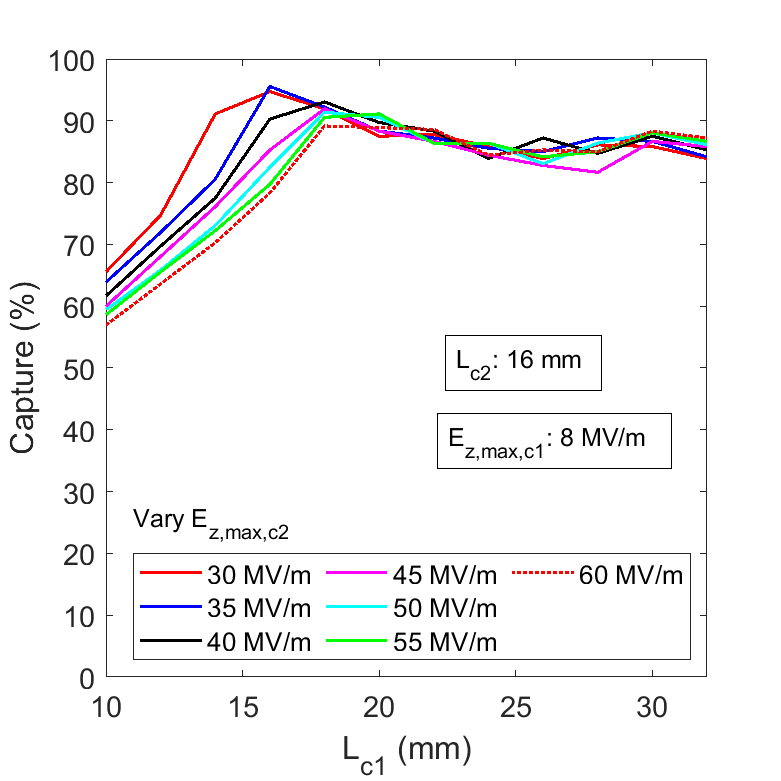}}{0.145in}{1.50in}	\\
		\end{tabular}
	\caption{1D code coarse scan results.}
	\label{fig:1DScan1}
\end{figure*}

The highest capture of 95.6$\%$ is achieved and the parameters are given in the first column of Table~\ref{tab:1DScans}. The part of the scans around the highest capture are plotted in Fig.~\ref{fig:1DScan1}. In each sub-figure, two variables are varied while the other two parameters are set to the optimum. The capture is more sensitive to the first cells' parameters than the second cells'. It is least sensitive to E$_{z,max,c2}$ and most sensitive to E$_{z,max,c1}$. This information is useful in guiding RF design as they provide parameter sensitivities and tolerances.  

\begin{table*}
	\caption{Selected representative 1D scan results.}
	\begin{ruledtabular}
		\begin{tabular}{lcccccccc}
		& \multicolumn{4}{c}{Scan parameters}  	& \multicolumn{4}{c}{Results}  \\
	\cline{2-5} 	\cline{6-9}	&	 L$_{c1}$ &	E$_{z,max,c1}$ & L$_{c2}$ & E$_{z,max,c2}$ 	&  Capture & KE$_{ave}$ & $\sigma_{\textnormal{KE}}$ & KE$_{max}$ \\
		&	 (mm) &	(MV/m) & (mm) & (MV/m)	& ($\%$) & (MeV) & (MeV) & (MeV)\\
			 \hline
		Highest-capture	&	16 & 8 & 14	& 35 & 95.6 & 4.4 & 1.2 & 5.4 \\
		Highest-KE$_{ave}$	& 32 & 30 & 38 & 60 & 56.9 & 5.9 &  1.1 & 6.7 \\
		High-capture and high-KE$_{ave}$ &	19.2 & 7.4 & 30	& 54 & 90.8 & 5.2 & 1.3 & 6.1 \\
		\end{tabular}
	\end{ruledtabular}
	\label{tab:1DScans}
\end{table*}

In the optimization process, one has to keep in mind that the goal of the optimization is not solely to achieve the highest capture, but rather to achieve high capture with a compact structure. In other words, the capture needs to be reasonably high and the linac can not be long. To keep the linac compact and simple, we would like to limit the number of cells to 6. The beam should have an energy peak around 6~MeV, so we need at least 4 accelerating cells. Consequently, we can only use 2 cells for capturing and bunching. 

\begin{figure}
	\begin{tabular}{cc}
		\def\stackalignment{l}
		\topinset{\bfseries(a)}{\includegraphics[width=1.6in]{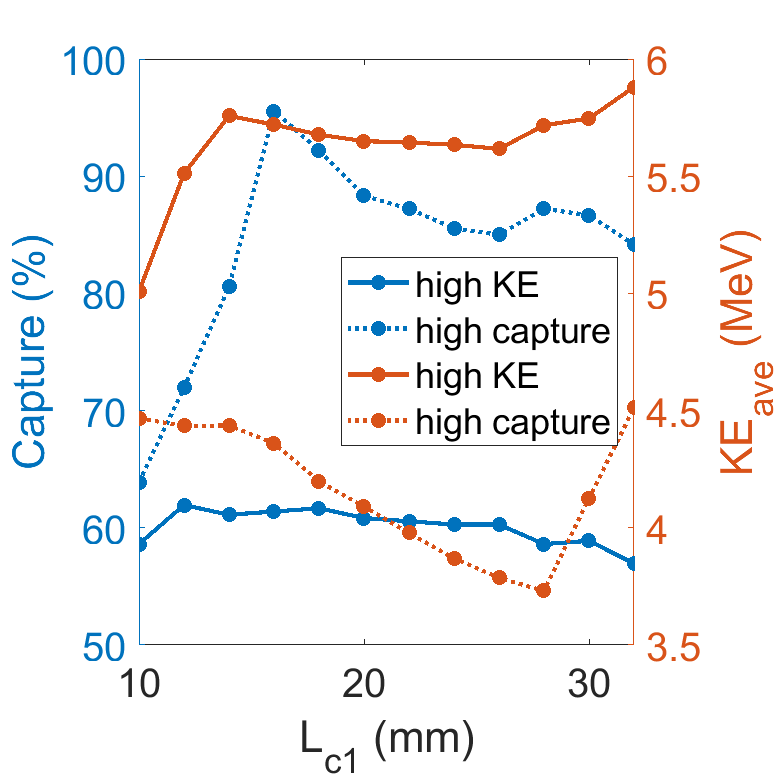}}{0.14in}{0.68in} 
		\def\stackalignment{l} 	&
		\topinset{\bfseries(b)}{\includegraphics[width=1.6in]{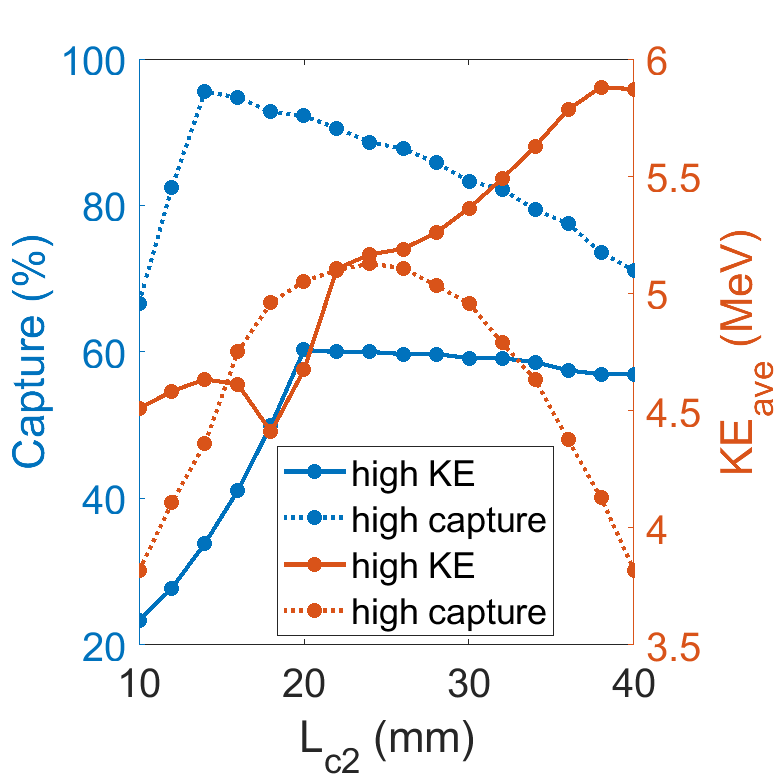}}{0.14in}{0.08in}	\\
		\topinset{\bfseries(c)}{\includegraphics[width=1.6in]{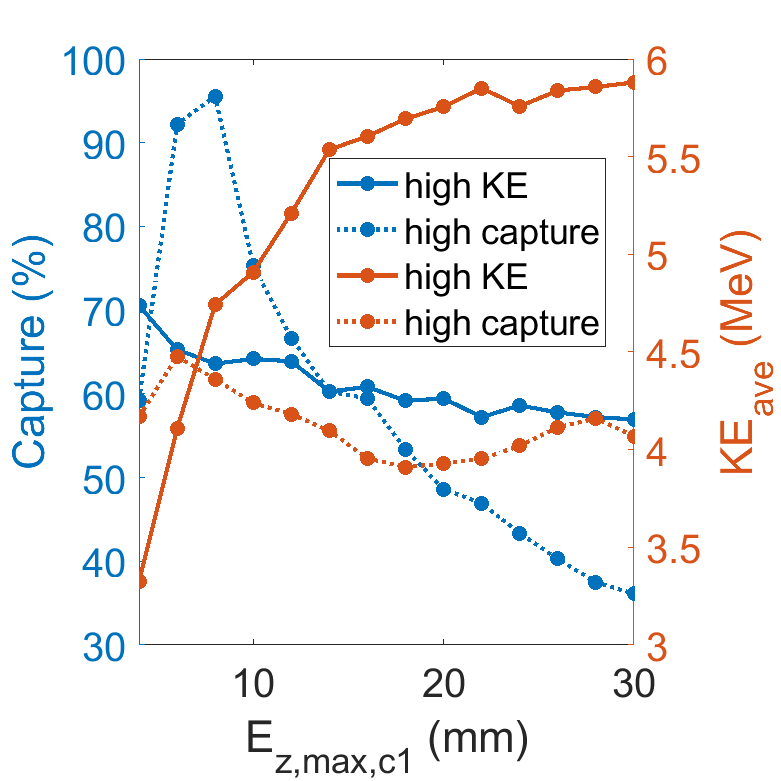}}{0.14in}{-0.18in} 
		\def\stackalignment{l} &
		\topinset{\bfseries(d)}{\includegraphics[width=1.6in]{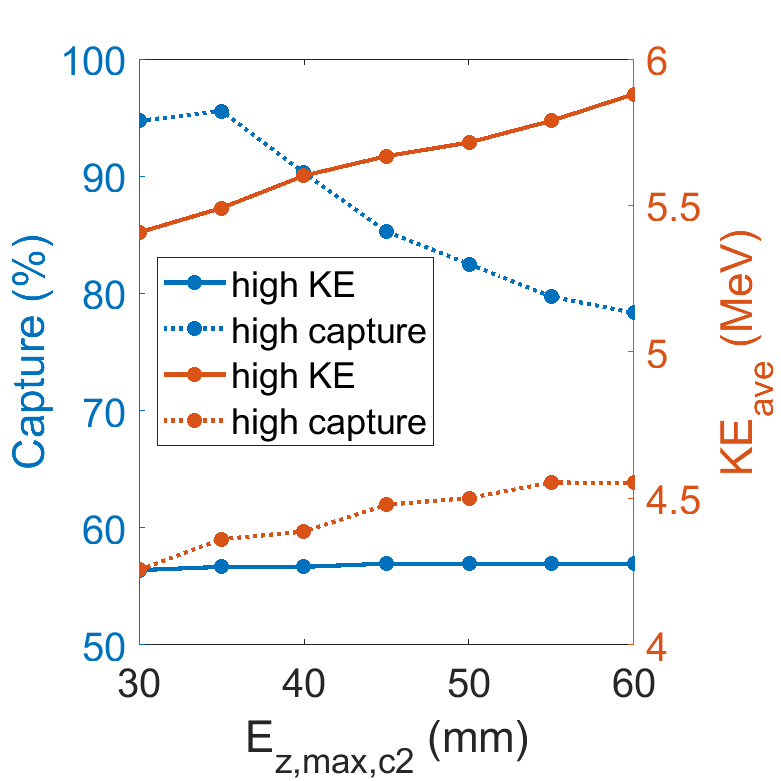}}{0.14in}{-0.18in}		\\
		\end{tabular}
	\caption{1D code coarse scan results around highest-capture and highest-KE$_{ave}$.}
	\label{fig:1DScan2}
\end{figure}

\begin{figure}
	{\scalebox{0.4} [0.4]{\includegraphics{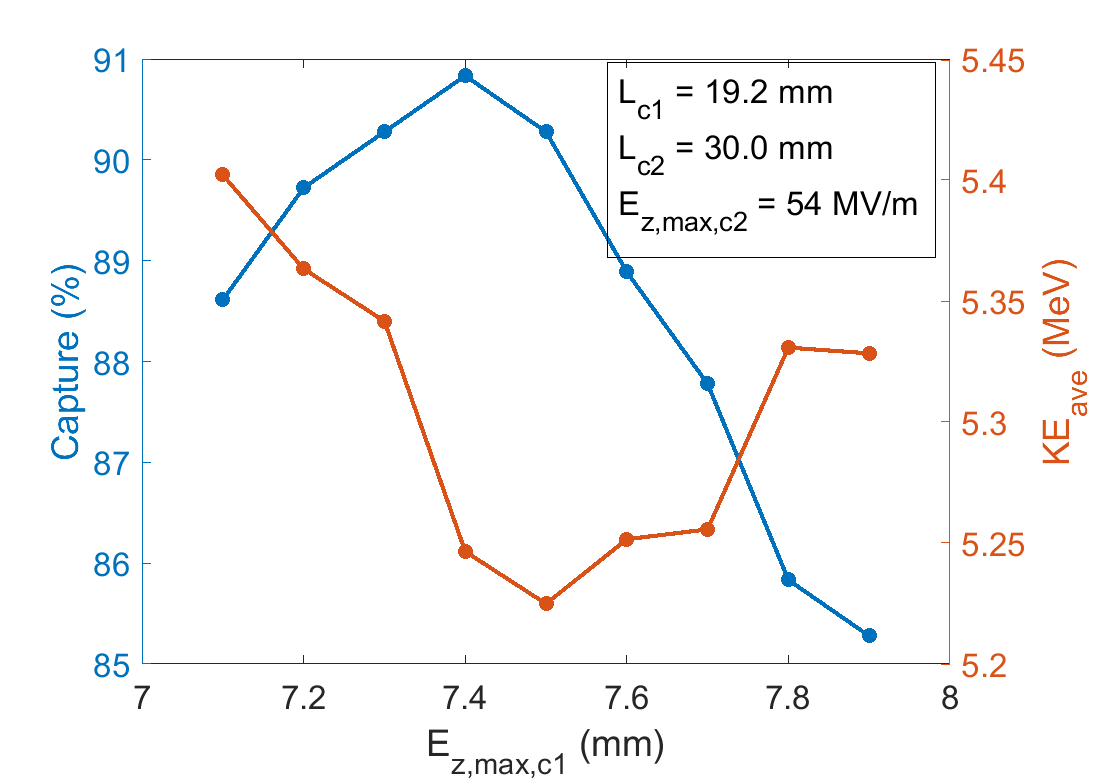}}}
	\caption{1D code fine scan results around high-capture and high-KE$_{ave}$.}
	\label{fig:1DScan3}
\end{figure}

Selected representative 1D scan results are given in Table~\ref{tab:1DScans}. While the highest-capture case has a capture of 95.6$\%$, the average KE KE$_{ave}$ is only 4.4~MeV and one would need to add extra cells to reach around 6~MeV. The highest-KE$_{ave}$ has KE$_{ave} = 5.9$, but the capture is only 56.9$\%$. Scan results around these highest-capture and highest-KE$_{ave}$ are shown in Fig.~\ref{fig:1DScan2}. It shows we cannot have the highest-capture and highest-KE$_{ave}$ at the same time. However, we can achieve reasonable high-capture ($>85\%$) and high-KE$_{ave}$ ($>$~5.2~MeV) at the same, as given the Table~\ref{tab:1DScans} (third and fourth columns) and shown in Fig.~\ref{fig:1DScan3}.

\begin{figure}
	{\scalebox{0.45} [0.45]{\includegraphics{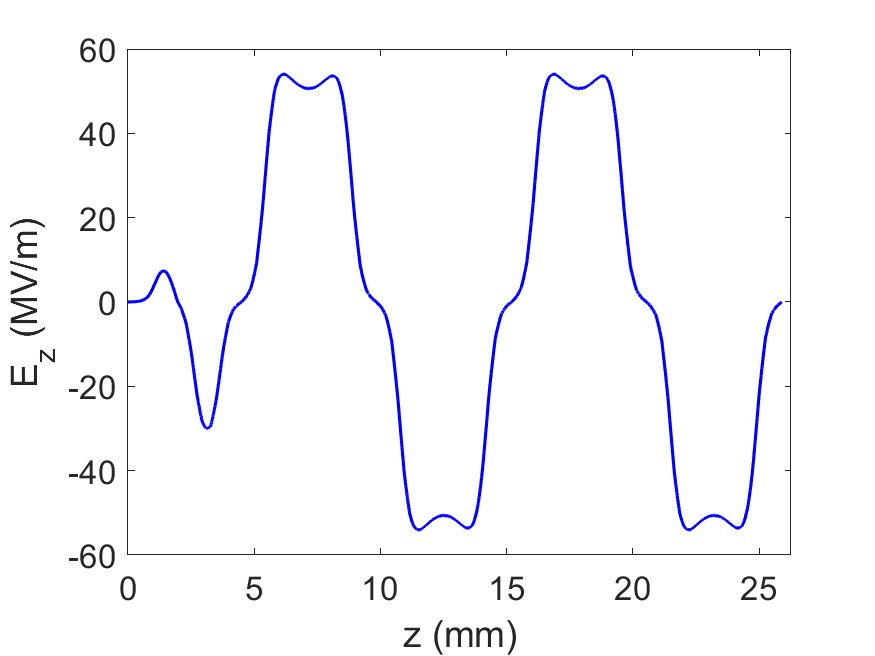}}}
	\caption{Optimized field profile from the 1D tracking code optimization.}
	\label{fig:1DOptEz}
\end{figure}

\begin{figure}
	{\scalebox{0.5} [0.5]{\includegraphics{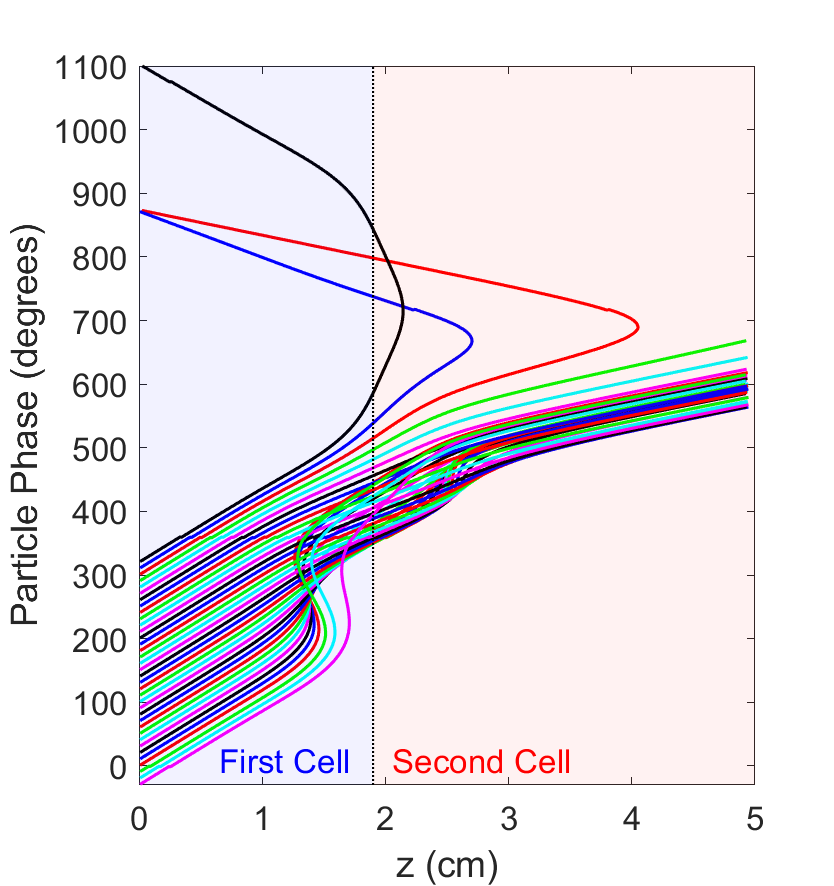}}}
	\caption{Applegate diagram generated with the optimized field profile of Fig.~\ref{fig:1DOptEz}}
	\label{fig:AppleGateOptim}
\end{figure}

\begin{figure}
	{\scalebox{0.45} [0.45]{\includegraphics{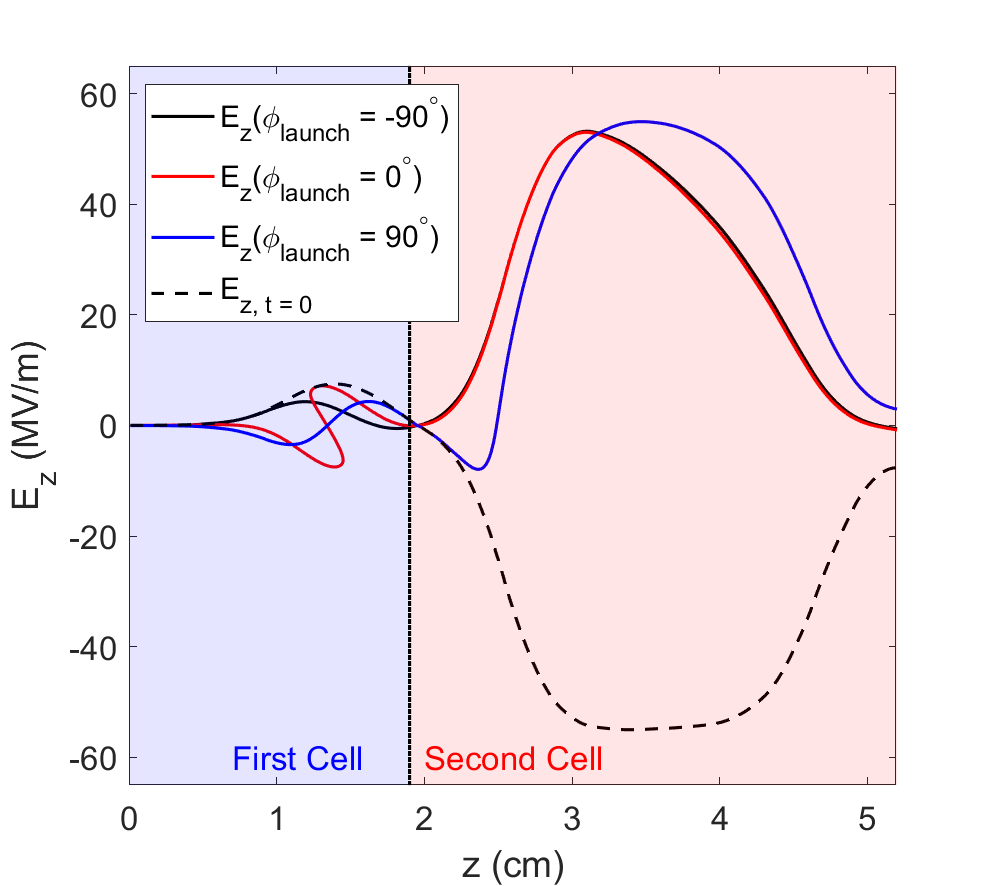}}}
	\caption{Recapture of the backstreaming electrons in the first cell. The dashed black line is the optimized field profile. The solid lines are the fields seen by the electrons launched at phases indicated in the parentheses.}
	\label{fig:EzvsZ}
\end{figure}

We used the optimized results in high-capture and high-KE$_{ave}$ case to generate the optimized field profile shown in Fig.~\ref{fig:1DOptEz}. The Applegate diagram shown in Fig.~\ref{fig:AppleGateOptim} and $E_{z}$ vs $z$ relations shown in Fig.~\ref{fig:EzvsZ} are generated by using the optimized field profile given. In Fig.~\ref{fig:AppleGateOptim}, we see that capture is increased significantly over that of the standard commercial linac shown in Fig.~\ref{fig:TradMedLinac}. 320$^\circ$ out of 360$^\circ$ phases are captured. The beam is well-bunched as well. Bunching begins in the first cell and continues in the second cell until at the exit of the second cell 320$^\circ$ of the launch phases are compressed to 105$^\circ$ at the exit. This is a threefold compression. 

The most novel aspect of using this field profile is the recapture of all the backstreaming electrons in the first cell, as shown in Fig.~\ref{fig:AppleGateOptim} and Fig.~\ref{fig:EzvsZ}. As the field amplitude is low in the first cell, backstreaming electrons were able to catch the next RF cycle and get accelerated. The only losses are from a few backstreaming electrons from the second cell. As the second cell has a high field amplitude, these backstreaming electrons have a larger KE, and thus cannot be recaptured. 

\begin{figure}
	\begin{tabular}{cc}
		\def\stackalignment{l}
		\topinset{\bfseries(a)}{\includegraphics[width=3in]{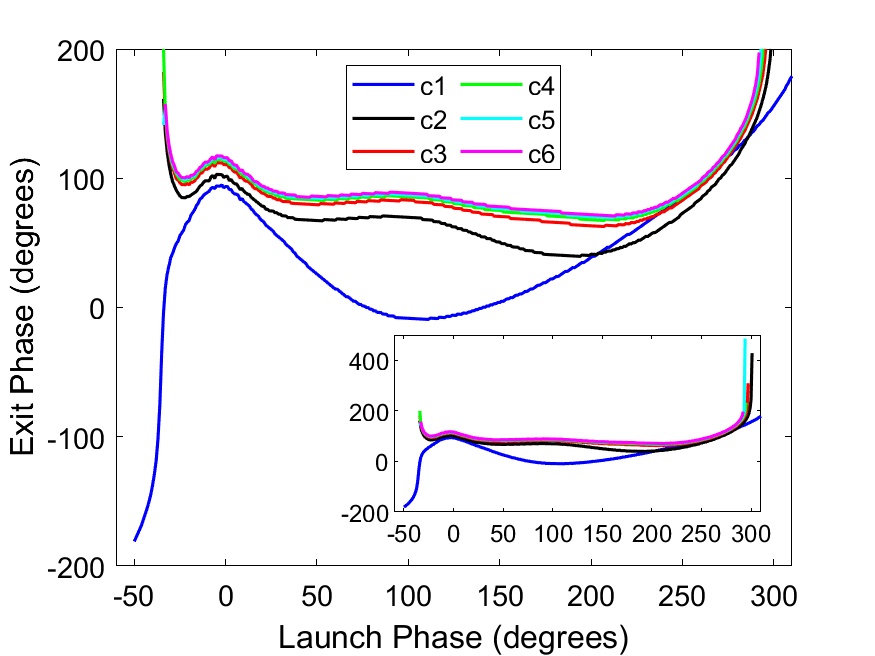}}{0.65in}{2.5in} \\
		\def\stackalignment{l}
		\topinset{\bfseries(b)}{\includegraphics[width=3in]{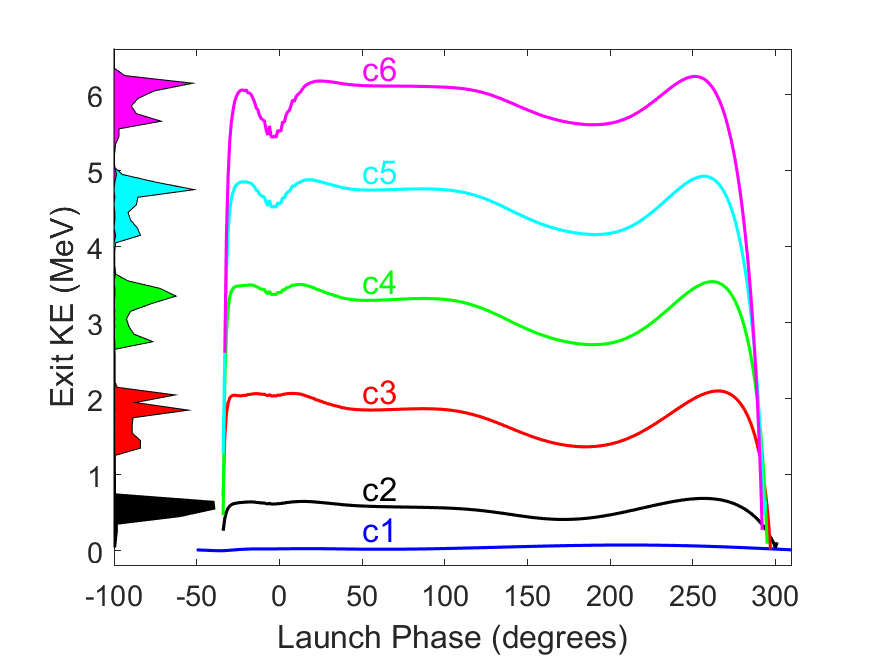}}{0.25in}{2.5in}		
	\end{tabular}
	\caption{1D tracking results with optimized field profile: (a) exit phase vs launch phase; and (b) exit KE vs launch phase. The inset of (a) is given to show a larger range of exit phases. The y-projection of the curves shows the KE spectrum (in arbitrary units) at the corresponding cell exits.}
	\label{fig:1DOpt}
\end{figure}

In addition to increasing the capture efficiency, the output energy spread is also reduced as a result of the improved bunching. The electrons are tracked through the optimized field shown in Fig.~\ref{fig:1DOptEz} and results are given in Fig.~\ref{fig:1DOpt}. The maximum of the $E_{z}$ of all the cells is set to 54~MV/m, except for the first cell, which is set to 7.5~MV/m. The lengths of the first and second cells are 19 and 33~mm, respectively, and other cells are 50~mm long. The captured phases are 325$^\circ$ out of the 360$^\circ$ launch phases - which gives a capture efficiency of about 90$\%$. A simulation performed with more electrons yielded a more precise estimate of 92.5$\%$ capture. As can be seen, only a few electrons launched at early and late phases are lost. The bunching is greatly improved compared to the existing commercial linac, as most of the exit phases after the third cell are within the $60^\circ$ to $120^\circ$ range. Fig.~\ref{fig:1DOpt} (b) shows that the exit KE spread is smaller compared to the existing linac. It is also shown that there is little acceleration gain in the first cell, rather it is used solely for bunching. 

\subsection{Verification of optimized field in ASTRA}
%7. Proving in ASTRA with generic sinusoidal field maps
The results of the 1D tracking code were then benchmarked using ASTRA and the results showed good agreement. In ASTRA simulations, 25~keV electrons were tracked through the optimized field profile shown in Fig.~\ref{fig:1DOptEz}. To create a 100~mA DC beam, electrons with a total charge of 0.1~nC are uniformly generated over a 1~ns time period, which covers 3 RF periods. The capture efficiency from the ASTRA simulations is 92$\%$, which is close to the results from using the 1D code. In these initial simulations, the cavity apertures are not included because we want to compare and verify the results of the 1D tracking code with ASTRA. Further simulations using ASTRA including the apertures are described in later sections. 

\begin{figure}
	{\scalebox{0.5} [0.5]{\includegraphics{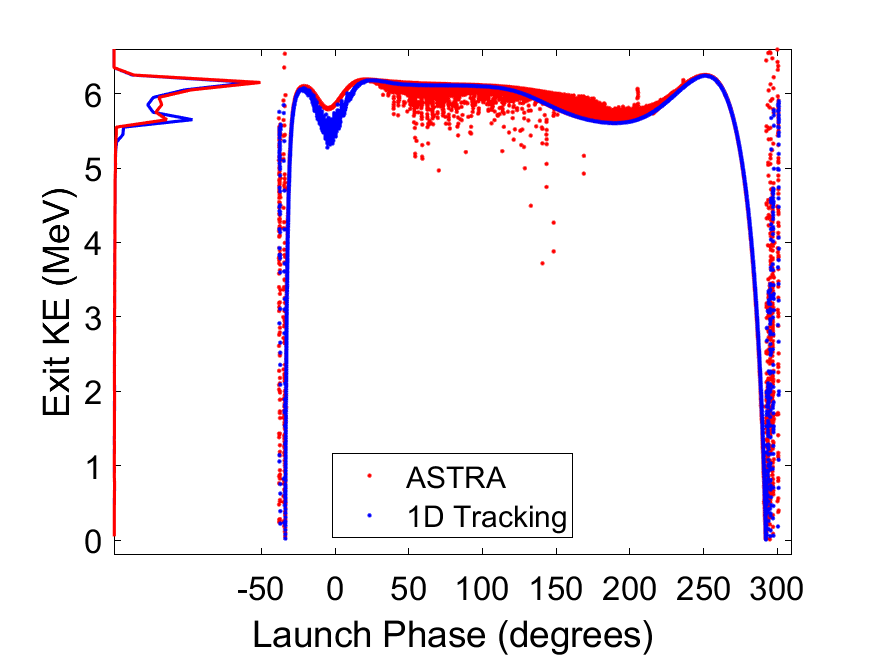}}} 
	\caption{Comparison of 1D tracking code and ASTRA tracking results at the exit of the linac. The solid lines are KE spectrum and the dots are KE as a function of launch phases.}
	\label{fig:KE1DvsASTRA}
\end{figure}

The exit KE spectrum and KE as a function of the launch phase, obtained by simulations in ASTRA and the 1D tracking code are shown in Fig.~\ref{fig:KE1DvsASTRA}. As can be seen, the two codes agree well.  ASTRA simulations were performed both with and without space-charge, and no significant difference was observed, thus indicating that space charge is not dominant. Both codes captured and rejected similar launch phases and produced a similar KE spectrum. 

\subsection{Fine-tuning of linac parameters in ASTRA}
\begin{figure}
	\begin{tabular}{cc}
		\def\stackalignment{l}
		\topinset{\bfseries(a)}{\includegraphics[width=1.63in]{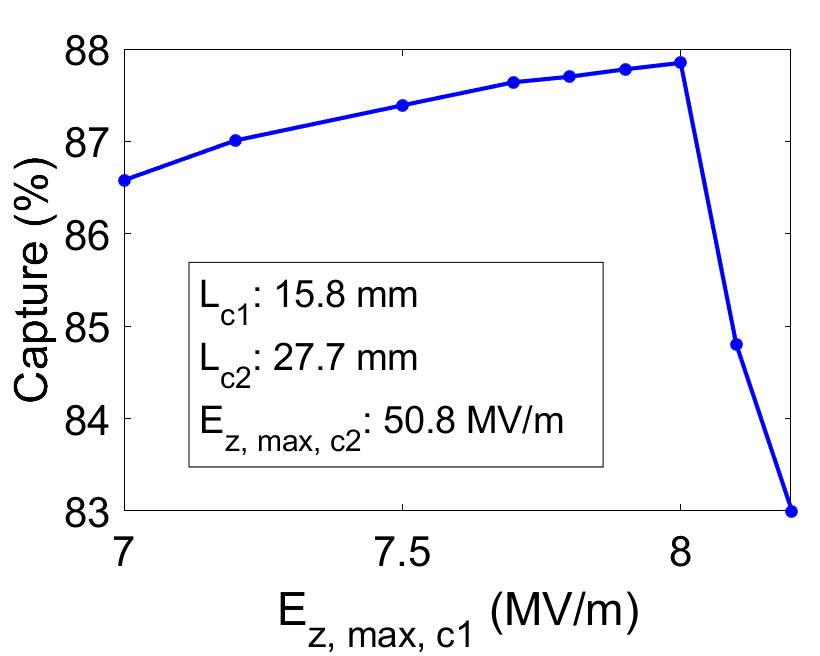}}{0.14in}{0.28in} 
		\def\stackalignment{l} &
		\topinset{\bfseries(b)}{\includegraphics[width=1.63in]{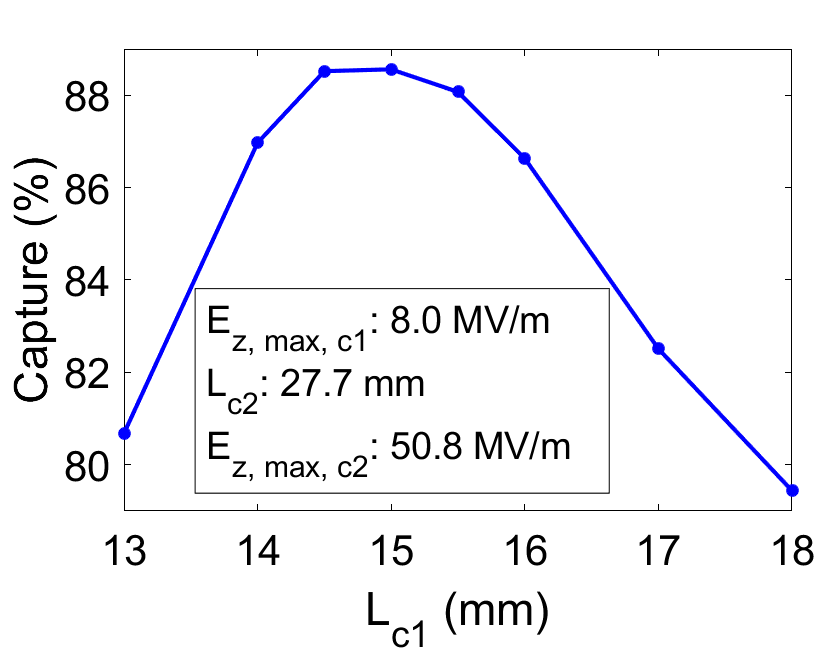}}{0.14in}{0.58in}		\\
		\topinset{\bfseries(c)}{\includegraphics[width=1.63in]{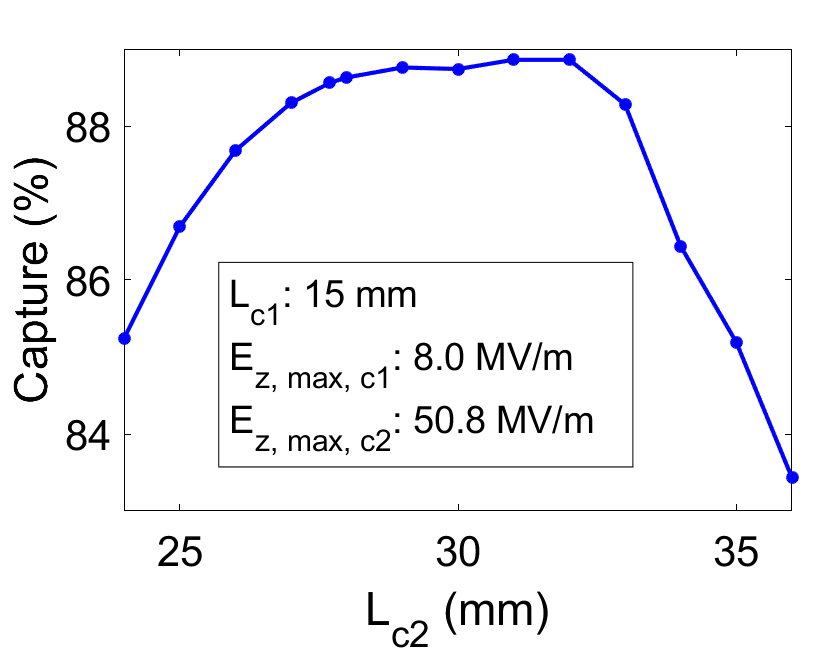}}{0.14in}{-0.44in} 
		\def\stackalignment{l} &
		\topinset{\bfseries(d)}{\includegraphics[width=1.63in]{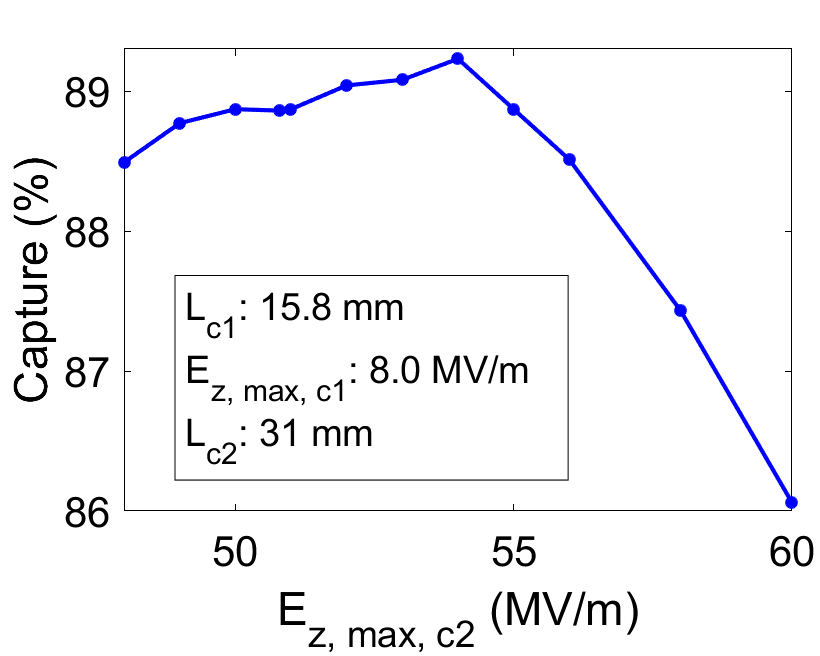}}{0.14in}{0.58in}		\\
		\end{tabular}
	\caption{Results of fine-tuning the cavity parameters using ASTRA.}
	\label{fig:FineTune}
\end{figure}

% In section IV I might add back in some of the parameters scans in ASTRA that tweaked the design up to the highest efficiency, showing the sensitivity to field amplitude in the first cell and length of the first two cells we can say it was done iteratively in CST.
The next step is fine-tuning the linac parameters in ASTRA. As with the optimization using the 1D code, the four parameters scanned are (a) the field amplitude of the first cell $E_{z, max, c1}$; (b) the length of the first cell $L_{c1}$; (c) the length of the second cell $L_{c2}$; and (d) the field amplitude of the second cell $E_{z, max, c2}$. 
Based on the global optimal results of 1D code, the new scans are conducted in much smaller steps for fine-tuning. As changing later parameters would have an impact on the earlier parameters, multiple rounds of scans are performed, where the new scans were performed with optimum results of earlier scans to make sure the final results are optimum for all the parameters simultaneously. 
The results are shown in Fig.~\ref{fig:FineTune}. The optimal values were found to be $L_{c1} = 15$~mm, $L_{c2} = 31$~mm, $E_{z, max, c1} = 8$~MV/m, and  $E_{z, max, c2} = 54$~MV/m. The amount of capture was found to be more sensitive to the parameters of the first cell than of the second cell. The capture decreases strongly when $E_{z, max, c1}$ is greater than 8~MV/m.

Note that in earlier simulations, the longitudinal position of the initial beam was at the inner wall of the cavity (noted as $z = 0$), which is not realistic. In an actual cavity, there needs to be at least a 1~cm distance between cavity entrance and initial beam position to account for the cavity wall and cathode flange thicknesses. Therefore, in the fine-tuning process, we shifted the initial beam position from $z=0$ to $z=-1$~cm and it caused a slight decrease of the capture to 87.4$\%$. However, the fine-tuning improved this to over 89$\%$. It can, however, be improved further to over 90$\%$ by optimizing the initial beam parameter as we will show in Section~\ref{OptimizingInitialBeam}.

\section{RF cavity design}
\begin{figure}
	{\scalebox{0.28} [0.3]{\includegraphics{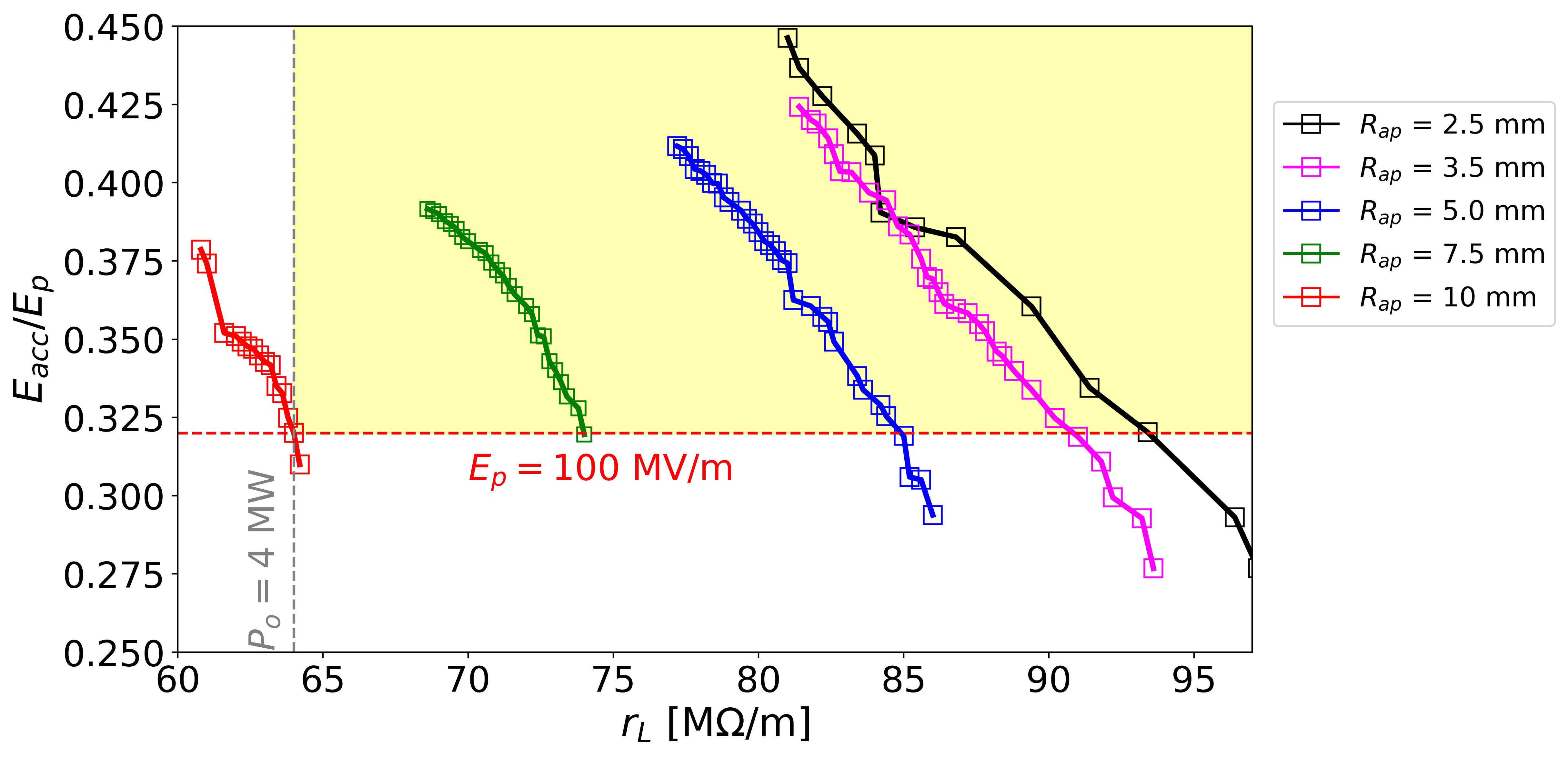}}}
	\caption{Pareto front for a middle cell of an periodic structure as a function of aperture radius}
	\label{fig:ParetoITAR}
\end{figure}

In order to keep manufacturing tolerances achievable while keeping the shunt impedance high, we have chosen a re-entrant side coupled $\pi/2$ mode cavity geometry. Due to the complexity of the behavior for some of the figures-of-merit in these multidimensional parameter spaces, we concluded that the use of Multi-Objective Genetic Algorithms (MOGA) optimization and the Pareto plots \cite{luo2019rf} to select the best combinations from a large set of variations of the geometry was the best approach. Fig.~\ref{fig:ParetoITAR} presents the results of a large number of simulations performed in the vast parameter space, highlighting the benefits of using the Pareto fronts. As previously discussed we require an aperture radius of 5~mm for high beam capture. If we scale the results to a 6 cell cavity capable of achieving 6~MeV accelerating voltage with a 4~MW input power while constraining the peak surface electric field to 100~MV/m we obtain a narrow parameter space of geometries that meet our requirements. We have chosen the furthest acceptable point to the right to minimize the required input power, giving a shunt impedance of just under 85~M$\Omega$/m.

Our coupling cells have a racetrack cross-section to minimize the transverse size while keeping the capacitive gap large enough (11~mm) to reduce the sensitivity to mechanical tolerances. The slot geometry was optimized to provide a 3.5~\% coupling factor, to reduce sensitivity to tolerances, while keeping to peak magnetic flux density to 214~mT at the design gradient.

The 1D cavity study showed that we need to have a short first cell with a low accelerating field which is difficult to realize in a real cavity. Studying the field minimization is complex as it is difficult to work out the frequency of each cell individually with side coupled cells and accelerating cells. To study minimizing the field in one cell while maintaining field flatness in the other cells, we performed a study of a 5 cell cavity where the coupling cells are identical to the main cells. We initially start with the cavity tuned for field flatness then modify individual cells. First, we studied varying the radius of the first (accelerating) cell and the second (coupling) cell. Varying only the first cell does indeed vary the field in that cell but at the expense of creating a finite field amplitude in the coupling cells which would be problematic in a real cavity due to the risk of multipactor and a need to minimize peak fields in the compact side coupling cavity shape. Varying only the second cell has no effect as there are no fields in the coupling cells if all other cells are tuned correctly, hence we next investigated varying the frequency of both the first and second cells. 
This small detuning of the first cell then allows the amplitude in the first cell to be varied by changing the frequency of the second cell. While this does indeed minimize the field in the coupling cells it was found to be impossible to have a lower field in the first cell without some field in the second cell. 
The reason for this is down to the reason that a $\pi/2$ has zero fields in the coupling cells. The accelerating cells on either side of a side coupling cell couple fields into the coupling cell that are 180 degrees different in phase and hence cancel each other out. To have zero fields in a coupling cell with a different field amplitude in the cells, either side requires that the coupling between each cell is modified inversely proportional to the field amplitude. We achieve this in our model cavity by having a larger aperture between cells 1 and 2 than the aperture between cells 2 and 3. It is found that this indeed does allow us to vary the field in the first cell to any amplitude without inducing a field in the coupling cells as long as the aperture is varied to match the coupling.

Our next issue is the need to have a short field in the first cell. Initially, we can reduce the cell length and the nose-cone gap to shorten the field profile. However, it is found this is limited to around 10~mm with a 5~mm aperture radius. Minimizing the gap also has the added advantage of reducing the accelerating field in that cell. This is due to the decay of the evanescent field in the input beam-pipe, hence to minimize the length of the field profile we need to reduce the aperture of the input beam-pipe. We find to minimize the field profile to the length given in the 1D studies we require a 3~mm aperture radius on the input beam-pipe. This is not dependant on the aperture radius between the cell and the adjacent cell and hence this can be kept at 5~mm to minimize beam loss.

\begin{figure}
	\begin{tabular}{cc}
		\def\stackalignment{l}
		\topinset{\bfseries(a)}{\includegraphics[width=3.3in]{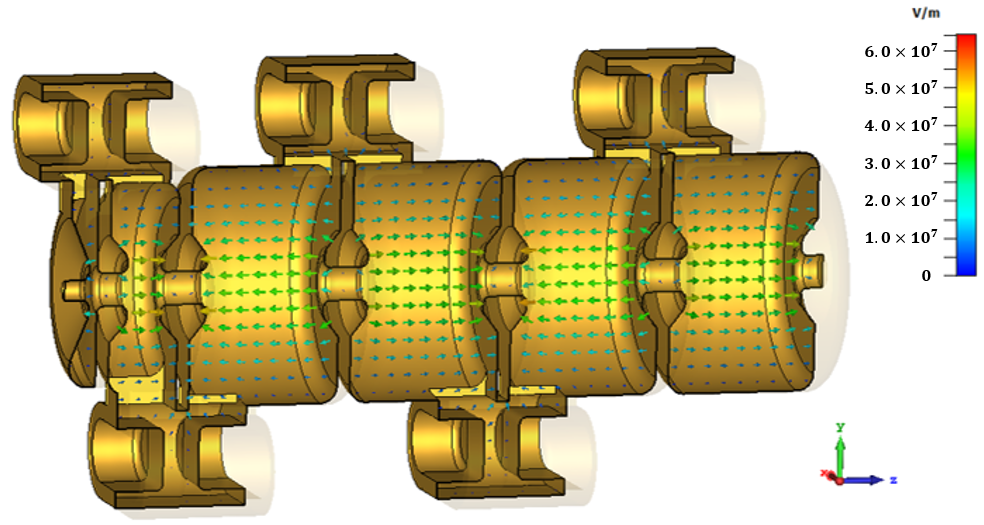}}{0.15in}{1.6in}\\
		\def\stackalignment{l}
		\topinset{\bfseries(b)}{\includegraphics[width=3.15in]{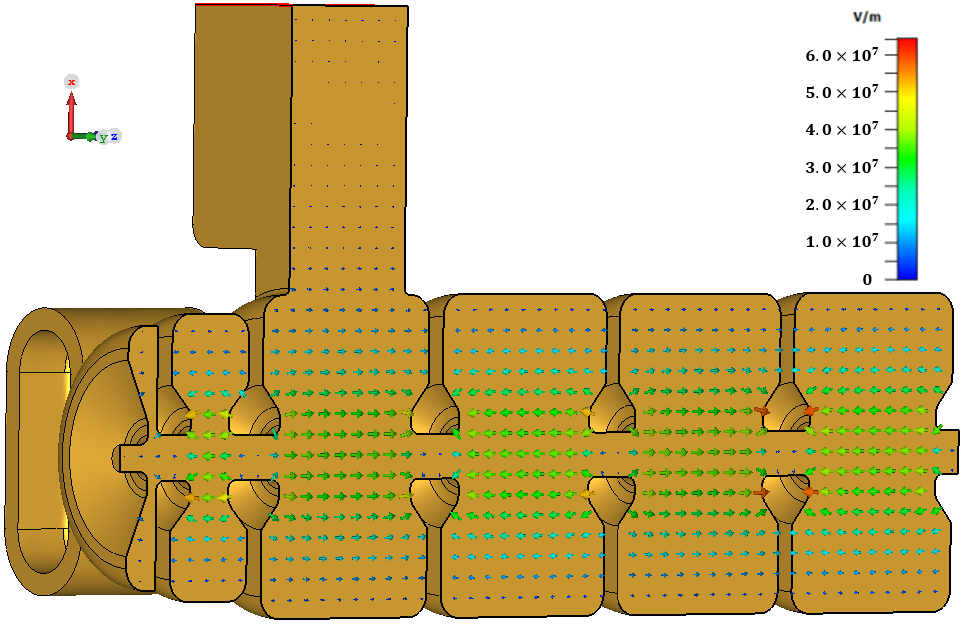}}{0.15in}{1.5in}	
	\end{tabular}
	\caption{The final cavity geometry showing the electric field map at 1~J of stored energy. Sub-figure (a) features side coupled cells and (b) features RF coupler.}
	\label{fig:ITARcavity}
\end{figure}

The results of these studies were translated into a side-coupled cavity with 6 accelerating cells and 5 coupling cells. The input beam pipe has an aperture radius of 3~mm while all other aperture radii are 5~mm. The coupling slots between the accelerating cells and coupling cells were each optimized to minimize the field in the coupling cells. This proved difficult in practice due to the complexity of the model, the need to re-tune both cells when the coupling slot is modified, and the fact that the side coupled cells are highly re-entrant and hence a small stored energy results in a large electric field. Despite this, it was possible to minimize the field in the coupling cells to 6.5~MV/m at the design gradient. Several iterations were performed with the ASTRA simulations to optimize the field profile, as described in the following section. For the final design, we require a peak input power of 2.65~MW to reach the design goal of 54~MV/m maximum on-axis electric field. This gives peak surface fields of 89~MV/m and 214~mT for the electric and magnetic fields respectively, which are well below the maximum limits. The electric field in the final cavity design is shown in Fig.~\ref{fig:ITARcavity}.

To feed the RF power into the linac, a waveguide coupler was integrated to the third cell, shown in Fig.~\ref{fig:ITARcavity} (b). The 3rd cell was chosen as the field levels in the first two cells are critical and hence to reduce variation the coupler is placed close to these cells. It is envisioned that this RF coupler will have a side vacuum port (not drawn), which is similar to the setup of the Ref~\cite{Yaqub2021}. The vacuum port will split into two ports: one used for rough pumping and the other for an ion pump. The RF coupler will be separated from the waveguide by an RF vacuum window. 

\section{Beam dynamics simulations with realistic cavity model}
\begin{figure}
	{\scalebox{0.45} [0.45]{\includegraphics{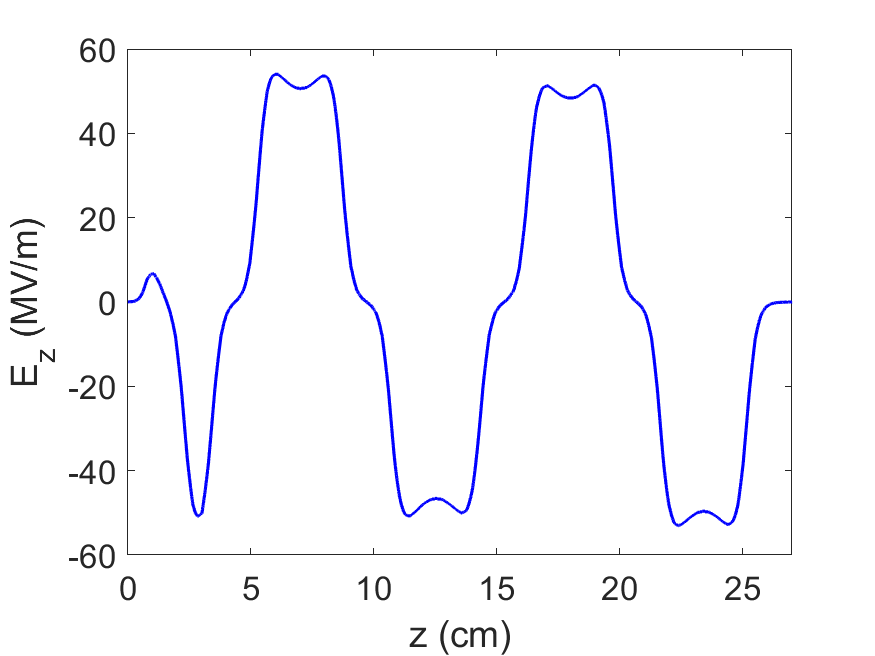}}}
	\caption{Field profile of entire cavity generated in CST simulation.}
	\label{fig:CSTEz}
\end{figure}

%9. 1D and ASTRA sims with a real cavity (a few different field maps from RF cavity with say different cell lengths)
Beam dynamics simulations of the realistic cavity model are implemented by taking the field profile of the entire cavity from CST, given by the red curve in Fig.~\ref{fig:CSTEz}, and using this as input into ASTRA. Apertures (cavity irises) and space charge are also included in the simulations. The field amplitude is scaled up so the maximum longitudinal electrical field $E_{z,~max}$ is 54~MV/m. The field from CST has smaller field amplitudes in some cells than the idealized field profile, which lowers the final KE. Several iterations of slightly modifying the CST model and tracking in ASTRA were performed to optimize the final cell lengths and field amplitudes. The resultant cell lengths are mostly the same as they were for the 1D model. In the ASTRA simulations, the beam was started at $-1$~cm as the minimum required distance between cavity inner wall and cathode exit is $1$~cm. The linac aperture is 5~mm in diameter, and the iris thicknesses are also 5~mm. The initial beam longitudinal profile is created as a 1~ns long flat distribution with 0.1~nC charge, which is equivalent to 100~mA quasi DC beam covering more than 3 RF cycles. The other initial beam parameters are given in Table~\ref{tab:E2Vgun}.

\begin{table}
	\caption{Initial electron beam parameters.}
	\begin{ruledtabular}
		\begin{tabular}{lll}
			Parameter & Unit & Value \\
			\hline
			Beam current			& mA & 100\\
			Pulse length            & ns & 1\\
			Initial beam position   & mm & $-$10\\			
			RMS transverse beam size& mm & 0.2 \\
			correlated divergence   & mrad & $-$12 \\
			transverse emittance    & $\pi\cdot$mm$\cdot$mrad   & 0.158 \\
			Average KE              & keV       &   25 \\
			KE energy spread        & eV     &  6.7 \\
		\end{tabular}
	\end{ruledtabular}
	\label{tab:E2Vgun}
\end{table}

Due to the differences between the optimal field profile and the field profile simulated in CST, particularly in terms of field flatness, the capture efficiency is slightly less than optimal. Scaling the field profile from CST to a max E$_{z,max}$ of 54~MV/m, the capture is 86$\%$, with loss due to backstreaming of 9.4$\%$, and loss at apertures of 4.6$\%$. However, when the E$_{z,max}$ is increased to 64~MV/m, the capture is 90.5$\%$ as shown in Table.~\ref{tab:EzmaxScan}. For radiotherapy applications, we want to follow the standard for commercial linac operations and limit the maximum peak surface field, E$_{pk}$, to well below 100~MV/m to reduce the risk of breakdowns. In our cavity design, we estimate E$_{z,max}$/E$_{pk} = 0.607$. This gives E$_{pk}$ of 89~MV/m for E$_{z,max}$ of 54~MV/m, which is well below the industrial safe operation limit.

\begin{table}
	\caption{Capture efficiency for different E$_{z,max}$.}
	\begin{ruledtabular}
		\begin{tabular}{lccccc}
			E$_{z,max}$ (MV/m) & 54   & 55 & 58   &  64  &  66  \\
			\hline
			Capture ($\%$)  & 86.0 & 86.6 & 88.1  & 90.5 & 86.6 \\
			back-streaming loss ($\%$) 	&  9.4 & 9.1 & 8.4 & 7.1 & 9.3  \\
			aperture loss ($\%$) & 4.6 & 4.3 &  3.5 & 2.4 & 4.1
		\end{tabular}
	\end{ruledtabular}
	\label{tab:EzmaxScan}
\end{table}

\section{Optimizing initial beam parameters}
\label{OptimizingInitialBeam}
\begin{figure}
	\begin{tabular}{cc}
		\def\stackalignment{l}
		\topinset{\bfseries(a)}{\includegraphics[width=1.67in]{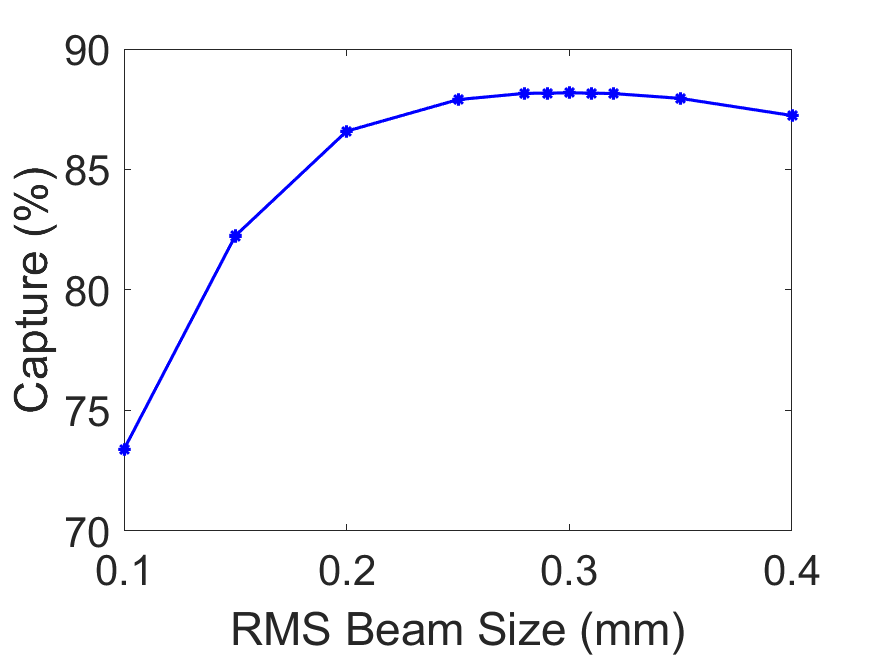}}{0.15in}{0.25in}&\topinset{\bfseries(b)}{\includegraphics[width=1.67in]{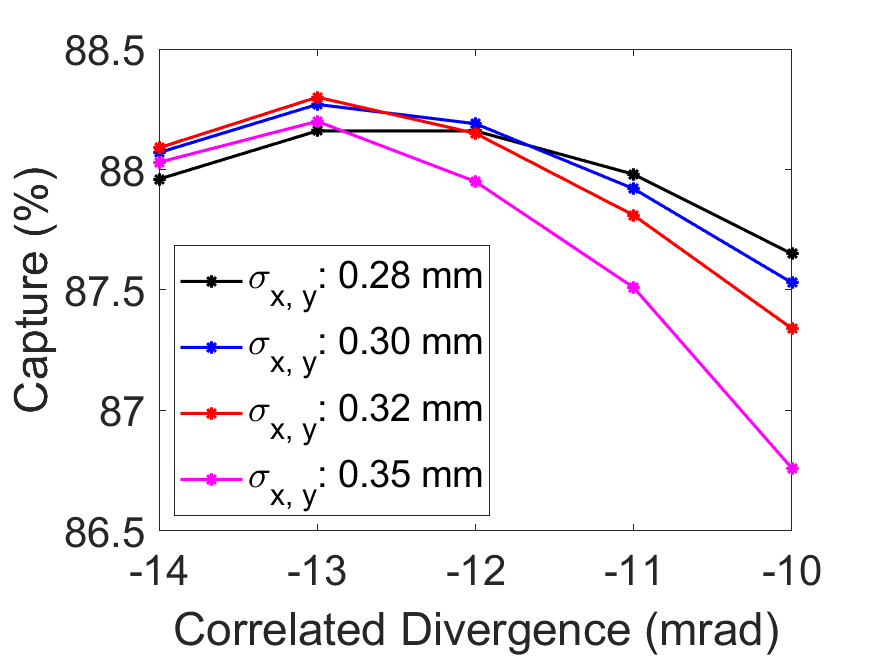}}{0.15in}{0.55in}\\
		\def\stackalignment{l}
		\topinset{\bfseries(c)}{\includegraphics[width=1.67in]{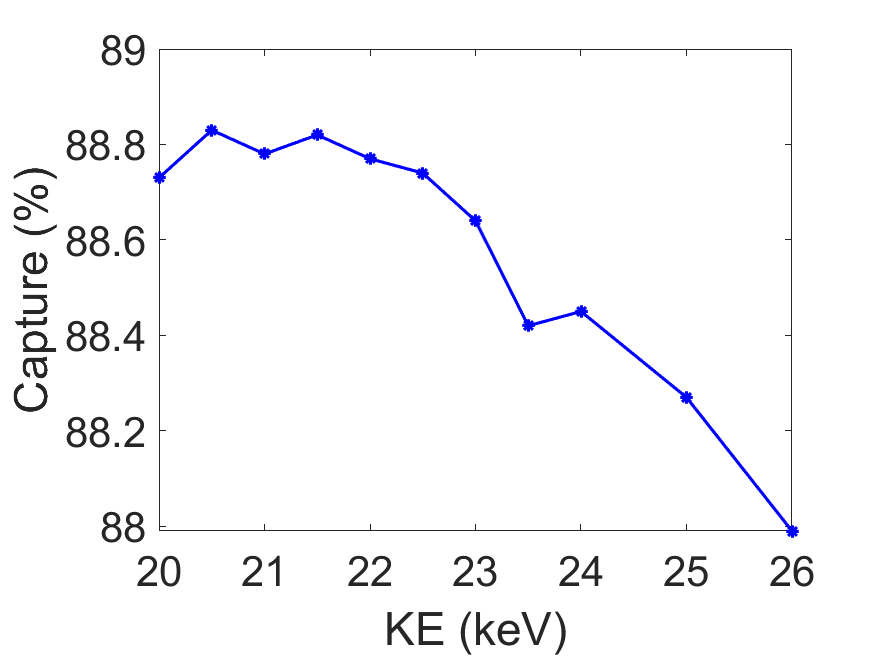}}{0.15in}{1.3in}&\topinset{\bfseries(d)}{\includegraphics[width=1.67in]{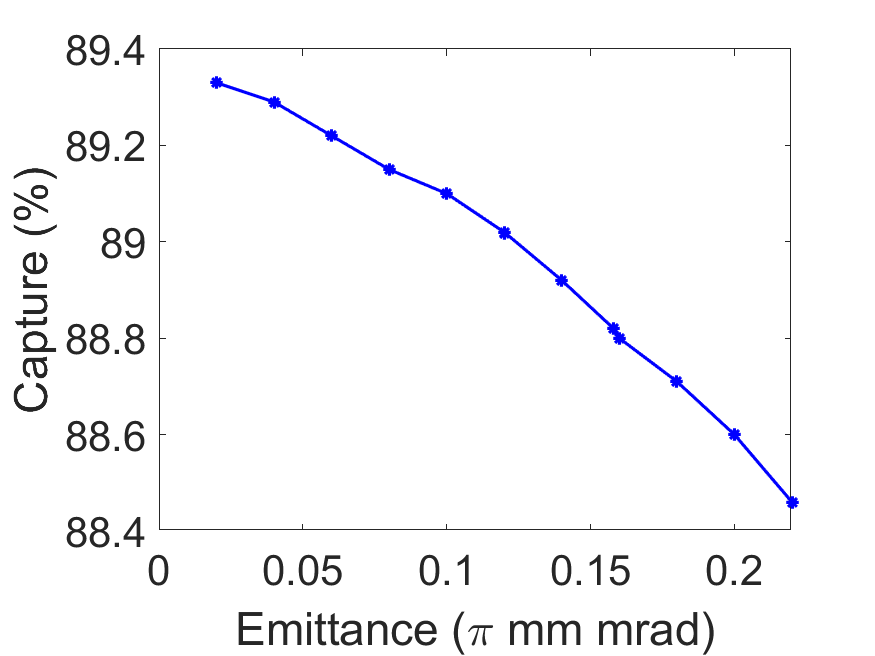}}{0.15in}{0.55in}	
	\end{tabular}
	\caption{ASTRA simulations showing scans of the initial beam parameters: (a) transverse RMS beam size; (b) correlated divergence; (c) kinetic energy; (d) emittance;}
	\label{fig:iniScan}
\end{figure}

%10. Possible something on trans. acceptance from the gun ???
So far we have optimized capture efficiency by varying the cavity parameters but used fixed initial electron beam parameters. However, the initial beam parameters can also be tuned to increase the amount of capture. Further ASTRA simulations were performed scanning the following parameters in order: transverse RMS beam size, beam divergence, KE, and emittance. The results are given in Fig.~\ref{fig:iniScan}. The highest capture increased from 86$\%$ to 89$\%$ when the initial RMS beam size $\sigma_{x, y}$ is 0.3~mm. The correlated divergence scan showed 0.3~mm and $-13$~mrad beam has the highest captures of 88.30$\%$. A lower KE of 21.5 keV pushed the capture to 88.82$\%$. Although the ASTRA scan shows a smaller emittance increases the capture, generating a beam with emittance smaller than 0.1~$\pi~mm~mrad$ may not be realistic. Simulations of a thermionic gun show that emittance of around 0.1~$\pi~mm~mrad$ should be achievable. Overall, if the initial beam parameters are optimized, we can increase the capture by $3\%$ to $89\%$. 

The capture can be increased further by combining the optimized initial beam parameters (0.3~mm and $-13$~mrad) with a higher E$_{z,max}$, as shown in Table.~\ref{tab:EzmaxScan2}. The emittance was set to a realistic achievable value of 0.158~$\pi\cdot$mm$\cdot$mrad. The scan was performed for beams with initial KEs of 21.5 and 25~keV. For commercial applications, safe operation is of paramount importance and the linac should be operated with lower E$_{z,max}$, in the range of 54$-$55~MV/m. This gives capture is around 89\% for an initial KE of 21.5~keV. Scientific research linacs, however, are often run at a higher gradient compared to commercial counterparts as there are sufficient monitoring, safety interlocks, and diagnostic systems, and the cavities manufactured with higher quality. Thus at the higher E$_{z,max}$ of 65~MV/m, a capture of 91.4\% can be achieved with a beam with an initial KE of 25~keV.

\begin{table}
	\caption{Capture efficiency for different E$_{z,max}$ with optimized initial beam parameters.}
	\begin{ruledtabular}
		\begin{tabular}{lcc}
			E$_{z,max}$ (MV/m) &	\multicolumn{2}{c}{Capture ($\%$)}   \\
			\cline{2-3}
			 		& {KE = 21.5~keV} &  {KE = 25~keV}  \\
			\hline
		 54  & 88.82 & 88.27 \\
		 55  & 89.23 & 88.67\\
		 58  & 90.26 & 89.76\\
		 64  & 84.32 & 91.37\\
		 66  & 82.18 & 87.60\\
		\end{tabular}
	\end{ruledtabular}
	\label{tab:EzmaxScan2}
\end{table}

\section{Discussion}
In Ref.~\cite{Kutsaev2021}, we see it is difficult to achieve compactness and high capture at the same time. While it is technically possible to achieve above 90$\%$ capture by using long bunchers ($>$10 cells) that are operated under low RF powers, the beam would experience other issues like space charge (as the beam is accelerated slowly) and RF defocusing effects, which would require external focusing. Besides, these long bunchers are low-$\beta$ structures, which are very low in RF power consumption efficiency. All these cons will render such cavities with long bunchers non-viable for commercial applications, and hence they are only used in scientific facilities. 
Liu \textit{et al.}~\cite{Liu2006} also designed an S-band traveling wave (TW) linac with 90$\%$ capture efficiency, but the cavity uses 59 cells over 2~m to reach 10~MeV energy. This corresponds to 5~MV/m gradient, which is less than $\frac{1}{4}^{\textnormal{th}}$ of design. Yurov \textit{et al.}~\cite{Yurov2017} were able to design a 1$-$3~MeV continuous-wave (CW) linac with a capture efficiency of 50$\%$, which is just slightly higher than existing industrial linacs.

We summarized the linacs and linac designs in the Ref.~\cite{Kutsaev2021} and compared them to our design as shown in Table~\ref{tab:CompareLinacs}. We find our linac design is not only the most compact one but has the highest capture as well. Unlike other high capture designs, our design doesn’t require extra components and does not add complexity to the system, yet the beam energy spread and size are kept small.

\begin{table*}
	\caption{Comparison of existing linacs and designs~\cite{Kutsaev2021} to novel high capture linac.}
	\begin{ruledtabular}
		\begin{tabular}{lcccc}
			Linacs & Mean/Peak Energy &  Length & Gradient & Capture  \\
			 & (MeV) & (m) & (MV/m) & (\%)  \\
			\hline
		 S-band with pre-buncher and 9.15-cm drift  & 6.8/NA &	0.7	& 9.7 &	77  \\
		 S-band TW option (a)  &  7.8/9.0	& 1.18 &	6.6 &	66 \\
		 S-band TW option (b)  &  9.0/9.3 &	1.25 &	7.2 &	65  \\
		 S-band SW side cuopled medical linac  & 6.5/7.4 &	0.33 &	19.7 & 30\\
		 X-band SW side cuopled medical linac  & 3.6/5.3 &	0.22 &	16.4 & 30\\
		 Ku-band split linac  & 0.14/0.18	 & 0.09 & 	1.6 & 	27  \\
		 C-band TW with amplitude = 2  &  4.5/6.0 & 	0.45 & 	10.0 & 	40  \\
		 C-band TW with amplitude = 3  &  5.3/6.0 & 	0.32 & 	16.6 & 	43 \\
		 C-band TW with amplitude = 4  &  5.4/6.0 &    0.31 & 	17.4 & 	51  \\
		 Novel S-band SW high capture linac  &  6.0/6.4 & 0.27 & 	22.2  & 90  \\

		\end{tabular}
	\end{ruledtabular}
	\label{tab:CompareLinacs}
\end{table*}

\section{Conclusion}
Increasing the average lifetime of the key components of commercial medical and security linacs is one of the main drivers for reducing linac downtime. One component known to regularly fail is the electron gun \cite{ClinicalOncology}. To address this we have designed an S-band thermionic gun-based linac with a high capture efficiency of 90$\%$, achieved using low energy velocity bunching with the inclusion of a low gradient short bunching cell at the entrance of the linac. As backstreaming and aperture losses are reduced by more than 80$\%$ compared to existing standard linacs, the back-bombardment effect is reduced, beam quality will be improved, and the cathode can be operated at a lower current, therefore the cathode lifetime will increase. Additionally, as there is less parasitic radiation, the linac will need less radiation shielding reducing the space around the linac on the gantry. The linac design can be easily implemented to existing commercial and research thermionic gun linacs without extra parts and cost, which in the long term can reduce operation and maintenance costs. This cavity design has a great potential to be implemented in commercial and research linacs to improve overall machine performance and lifetime. A new linac based on this design will be manufactured and tested in due course.

\begin{acknowledgments}
The following project was supported by STFC (GCRF) grants ST/S002081/1 and ST/S001190/1. The author would like to thank the staff at CERN and the International Cancer Expert Corps (ICEC) for support and advice during this project, and well as Taofeeq Ige, National Hospital Abuja, Hubert Foy, Africsis, and Surbhi Grover, Hospital of University of Pennsylvania for providing information on requirements for radiotherapy linacs in Africa and encouragement on this project.
\end{acknowledgments}

%\bibliography{bibtex}

%merlin.mbs apsrev4-1.bst 2010-07-25 4.21a (PWD, AO, DPC) hacked
%Control: key (0)
%Control: author (8) initials jnrlst
%Control: editor formatted (1) identically to author
%Control: production of article title (-1) disabled
%Control: page (0) single
%Control: year (1) truncated
%Control: production of eprint (0) enabled
%

%\bibliographystyle{ieeetr}
\end{document}